\documentclass[british,prb,superscriptaddress]{revtex4-1}
\usepackage[T1]{fontenc}
\usepackage[latin9]{inputenc}
\usepackage{geometry}
\usepackage{amsmath}
\usepackage{graphicx}
\usepackage{color}
\usepackage{wrapfig}

\makeatletter
\usepackage{babel}

\makeatother

\usepackage{babel}
\begin{document}

\title{Active Vertex Model for Cell-Resolution Description of Epithelial Tissue Mechanics}

%
%
%
\author{Daniel L. Barton}

\affiliation{Division of Physics, School of Science and Engineering, University
of Dundee, Dundee DD1 4HN, United Kingdom}
\affiliation{Division of Computational Biology, School of Life Sciences, University
of Dundee, Dundee DD1 4HN, United Kingdom}

\author{Silke Henkes}
\affiliation{Institute of Complex Systems and Mathematical Biology, Department
of Physics, University of Aberdeen, Aberdeen, AB24 3UE, United Kingdom}

\author{Cornelis J. Weijer}
\affiliation{Division of Cell and Developmental Biology, School of Life Sciences, University of Dundee, Dundee DD1 4HN, United Kingdom}

\author{Rastko Sknepnek}
\email[Correspondence should be addressed to: ]{r.sknepnek@dundee.ac.uk}
\affiliation{Division of Physics, School of Science and Engineering, University
of Dundee, Dundee DD1 4HN, United Kingdom}
\affiliation{Division of Computational Biology, School of Life Sciences, University
of Dundee, Dundee DD1 4HN, United Kingdom}

\begin{abstract}
We introduce an Active Vertex Model (AVM) for cell-resolution studies of the mechanics of confluent epithelial
 tissues consisting of tens of thousands of cells, with a level of detail inaccessible to similar methods. The
 AVM combines the Vertex Model for confluent epithelial tissues with active matter dynamics. This introduces a natural
 description of the cell motion and accounts for motion patterns observed on multiple scales. Furthermore, cell contacts
 are generated dynamically from positions of cell centres. This not only enables efficient numerical implementation,
 but provides a natural description of the T1 transition events responsible for local tissue rearrangements. The AVM
 also includes cell alignment, cell-specific mechanical properties, cell growth, division and apoptosis. In addition,
 the AVM introduces a flexible, dynamically changing boundary of the epithelial sheet allowing for studies of phenomena
 such as the fingering instability or wound healing. We illustrate these capabilities with a number of case studies.
\end{abstract}
\maketitle

\section{Introduction}

Collective cell migration\cite{ilina2009mechanisms,weijer2009collective}
in epithelial tissues is one of the key mechanisms behind many biological
processes, such as the development of an embryo,\cite{forgacs2005biological}
wound healing,\cite{martin1997wound,brugues2014forces} and tumour
metastasis and invasion.\cite{wirtz2011physics} Due to their layered,
tightly connected structure,\cite{guillot2013mechanics} epithelial
tissues also serve as an excellent model system to study cell migration
processes. Over several decades\cite{schatten2007multiscale} extensive
research efforts have been devoted to understanding molecular processes
that lead to cell migration\cite{lecuit2011force} and, at larger scales, on how cell
migration drives complex processes at the level of the entire tissue,
such as morphogenesis. With recent advances in various microscopy
techniques combined with the development of sophisticated automatic
cell tracking methods, it is now possible to study collective migration
patterns of a large number of cells over extended periods of time
with the cell-level resolution, both \emph{in vitro} and \emph{in
vivo}. Traction force microscopy\cite{harris1980silicone} experiments
revealed that collective cell motion is far more complex than
expected.\cite{angelini2011glass,tambe2011collective,trepat2011plithotaxis}
It is often useful to draw parallels between the collective behaviour
of tissues and systems studied in the physics of colloids, granular materials and foams as
these can provide powerful tools for understanding complex cell
interactions in biological systems. For example, in a homogenous cell
monolayer, one observes large spatial and temporal fluctuations of
inter-cellular forces that cannot be pinpointed to a specific location,
but cover regions that extend over several cells.\cite{sadati2013collective}
These are reminiscent of the fluctuations observed in supercooled
colloidal and molecular liquids approaching the glass transition\cite{angelini2011glass}
and include characteristic features of the dynamical and mechanical response, such as dynamical
heterogeneities and heterogeneous stress patterns, that were first observed in glasses,
colloids and granular materials and that have extensively been studied
in soft condensed matter physics.\cite{berthier2011dynamical} It
has also been argued that the migration patterns are sensitive to
the expression of different adhesion proteins\cite{bazellieres2015control}
as well as to the properties of the extracellular environment,\cite{charras2014physical}
such as the stiffness of the substrate.\cite{discher2005tissue,elosegui2014rigidity}
These observations lead to the development of the notion of \emph{plithotaxis},\cite{tambe2011collective}\emph{
}a universal mechanical principle akin to the more familiar chemotaxis,
which states that each cell tends to move in a way that maintains
minimal local intercellular shear stress. While plausible, it is yet
to be determined whether plithotaxis is indeed a generic feature in
all epithelial tissues.

Equally fascinating are the experiments on model systems that study
cell migration in settings designed to mimic wound healing.\cite{kim2013propulsion,brugues2014forces,das2015molecular,vedula2015mechanics,serra2015mapping}
For example, the existence of mechanical waves that span the entire
tissue and generate long-range cell-guidance have been established
in Madin-Darby Canine Kidney (MDCK) epithelial cell monolayers.\cite{serra2015mapping}
Subtle correlations between purse-string contractility and large-scale
remodelling of the tissue while closing circular gaps have also been identified.\cite{vedula2015mechanics}
Finally, a mechanism dubbed \emph{kenotaxis} has been proposed,\cite{kim2013propulsion}
which suggests that there is a robust tendency of a collection of
migrating cells to generate local tractions that systematically and
ooperatively pull towards the empty regions of the substrate.

On the developmental side, in pioneering work, Keller \emph{et al.}\cite{keller2008reconstruction}
constructed a light-sheet microscope that enabled them to track \emph{in
vivo} positions of each individual cell in a zebrafish embryo over
a period of 24h. A quantitive analysis\cite{manning2010coaction}
of the zebrafish embryo was also able to relate mechanical energy
and geometry to the shapes of the aggregate surface cells. Another
extensively studied system that allows detailed tracking of individual
cells is the\emph{ Drosophila} embryo.\cite{blanchard2009tissue,butler2009cell,keller2010fast,bosveld2012mechanical,collinet2015local}
In recent studies that combined experiments with advanced data analysis,
it was possible to quantitatively account for shape change of the
wing blade by decomposing it into cell divisions, cell rearrangements
and cell shape changes.\cite{etournay2015interplay,etournay2016tissueminer}
Finally, it has recently become possible to track more than 100,000
individual cells in a chick embryo over a period exceeding 24 hours.\cite{rozbicki2015myosin}
This was achieved by developing an advanced light-sheet microscope and
state-of-the-art data analysis techniques designed to automatically
track individual cells in a transgenic chick embryo line with the
cell membranes of all cells in the embryonic and extra embryonic tissues
labelled with a green fluorescent protein tag. All these experiments
and advanced data analysis techniques provide unprecedented insights
into the early stages of the embryonic development, making it possible
to connect processes at the level of individual cells with embryo-scale
collective cell motion patterns. 

While there have been great advances in our understanding of how cells
regulate force generation and transmission between each other and
with the extracellular matrix in order to control their shape and
cell-cell contacts,\cite{lecuit2011force} it is still not clear how
these processes are coordinated at the tissue-level to drive tissue
morphogenesis or allow the tissue to maintain its function once it reaches
homeostasis. Computer models of various levels of complexity have
played an essential role in helping to answer many of those questions.\cite{anderson2007single}
One of the early yet successful approaches has been based on
the cellular Potts model (CPM).\cite{graner1992simulation,glazier1993simulation}
In the CPM, cells are represented as groups of ``pixels'' of
a given type. Pixels are updated one at a time following a set
of probabilistic rules. Pixel updates in the CPM are computationally
inexpensive,\cite{izaguirre2004compucell} which allows for simulations
of large systems. In addition, the CPM extends naturally to three dimensions.\cite{swat2012multi} While very successful
in describing cell sorting as well as certain aspects of tumour growth,\cite{szabo2013cellular}
CPM has several limitations, the main one being that the dynamics
of pixel updates is somewhat artificial and very hard to relate to
the dynamics of actual cells. This problem has been to some extent
alleviated by the introduction of the subcellular element model (ScEM).\cite{newman2005modelling,sandersius2011emergent}
ScEM is an off-lattice model with each cell being represented as a
collection of 100-200 elements - spherical particles interacting with
their immediate neighbours via short-range soft-core potentials. Therefore,
ScEM is able to model cells of arbitrary shapes that grow, divide
and move in 3D. The main disadvantage of ScEM is that it is computationally
expensive (off-lattice methods in general require more computations
per time step compared to their lattice counterparts), and without
a highly optimised parallel implementation, applications of the ScEM
are limited to a few hundred cells at most, which is not sufficient
to study effects that span long length- and time-scales.

Particle-based models have also been very successful at capturing
many aspects of cell migration in tissues.\cite{drasdo2007role,zimmermann2014intercellular,zimmermann2016contact}
However, when it comes to modelling confluent epithelial layers with
the resolution of individual cells, one of the most widely and successfully
used models is the Vertex Model (VM). The VM originated
in studies of the physics of foams in the 1970s and was first applied
to model monolayer cell sheets in the early 1980s.\cite{honda1980much}
Over the past 35 years it has been implemented and extended numerous
times and used to study a wide variety of different systems.\cite{brodland2004computational,farhadifar2007influence,staple2010mechanics,fletcher2013implementing,fletcher2014vertex,Curran078204}
The VM is in the core of the cell-based\cite{pathmanathan2009computational}
CHASTE,\cite{mirams2013chaste} a versatile and widely used software
package. The VM assumes that all cells in the epithelium are roughly
the same height and thus that the entire system can be well approximated
as a two-dimensional sheet. The conformation of the tissue in the VM is computed
as a configuration that simultaneously optimises area and perimeter
of all cells. While the model is quasi-static in nature, it captures
remarkably well many properties of actual epithelial tissues.
There have been numerous attempts to introduce dynamics into the vertex
model.\cite{honda1980much,weliky1990mechanical,nagai2001dynamic,fletcher2013implementing,Curran078204}
However, there are limitations associated with each
approach. To the best of our knowledge, most dynamical versions
of the vertex model seem to neglect fluctuations with a notable recent
exception.\cite{Curran078204} In contrast, recent traction microscopy experiments\cite{sadati2013collective}
suggest that these fluctuations might be a crucial ingredient in understanding
collective cell migration. Finally, we point out a technical point
that makes the implementation of VM somewhat challenging. Namely,
in order to capture topology changing moves, such as cell neighbour exchanges, i.e., T1 transitions,
one has to perform rather complex mesh restructuring operations\cite{fletcher2013implementing,spencer2016vertex}
that require complex data structures and algorithms and that inevitably
add to the computational complexity of the model.

Building upon the recently introduced Self-Propelled Voronoi (SPV)
model,\cite{bi2016motility} in this paper we apply the ideas introduced
in the physics of active matter systems\cite{marchetti2013hydrodynamics}
to the VM. This allows us to construct a hybrid, Active Vertex Model
(AVM) that is able to accurately describe the collective migration dynamics of a large number of cells. The AVM is implemented within
\emph{SAMoS}, an off-lattice particle-based simulation software developed
specifically to study active matter systems. One of the key advantages
of the hybrid approach presented in this study is that it not only
enables studies of very large systems, but also introduces a very
natural way to handle the T1 transitions, thus removing the need for
complex mesh manipulations that are of uncertain physical
and biological meaning. 

The paper is organised as follows. In Section \ref{sec:model} we
briefly review the general features of the VM and derive expressions for
forces on individual cells central to the active vertex model. In
Section \ref{sec:examples} we apply the model to a number of test
cases. Finally, in Section \ref{sec:conclusions} we provide an outlook
for future applications of this model. Details of the force calculation
and implementation are presented in the Appendices. 

\section{Model}

\label{sec:model}

\subsection{Overview of the Vertex Model}

\label{subsec:overview}
\begin{wrapfigure}{r}{0.5\textwidth}
\begin{centering}
\includegraphics[width=0.5\columnwidth]{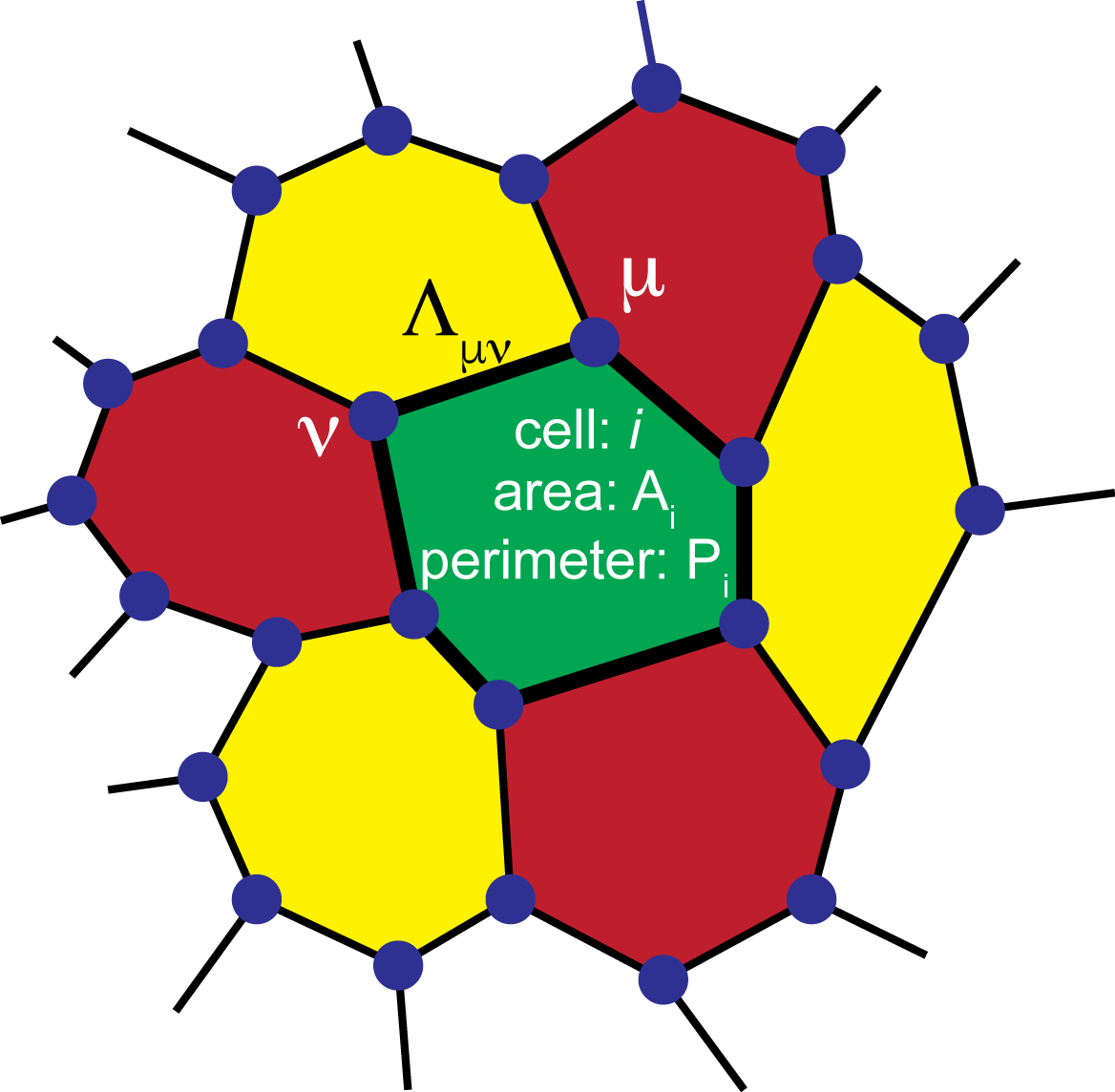} 
\par\end{centering}
\caption{In the Vertex Model (VM), a confluent epithelial sheet is represented
as a polygonal tiling of the plane with no holes or overlaps. Each
cell is represented by an $n-$sided polygon. Neighbouring cells share
an edge, which models the cell junction as a straight line. Three edges meet at a vertex (dark blue dots). The
behaviour of cell $i$ is described by three parameters: 1) reference
area $A_{i}^{0}$ , 2) area modulus $K_{i}$, and 3) perimeter
modulus $\Gamma_{i}$. In addition, a junction connecting vertices
$\mu$ and $\nu$ has tension $\Lambda_{\mu\nu}$. \label{fig:vertex_model}}
\end{wrapfigure}

Owing to its origins in the physics of foams,\cite{weaire2001physics}
in the VM cells are modelled as two-dimensional convex polygons that
cover the plane with no holes and overlaps, i.e., the epithelial tissue
is represented as a convex polygonal partitioning of the plane (Fig.
\ref{fig:vertex_model}). The main simplification compared to models
of foams is that the VM assumes that contacts between neighbouring
cells are straight lines. In addition, neighbouring
cells are also assumed to share a single edge, which is a simplification
compared to real tissues, where junctions between two neighbouring
cells consist of two separate cell membranes that can be independently
regulated. Typically, three junction lines meet at a vertex, although
vertices with a higher number of contacts are also possible.\cite{spencer2016vertex}
The model tissue is therefore a mesh consisting of polygons (i.e.,
cells), edges (i.e., cell junctions), and vertices (i.e., meeting points of three or more cells).

An energy is associated to each configuration of the mesh. It can
be written as
\begin{equation}
E_{VM}=\sum_{i=1}^{N}\frac{K_{i}}{2}\left(A_{i}-A_{i}^{0}\right)^{2}+\sum_{i=1}^{N}\frac{\Gamma_{i}}{2}P_{i}^{2}+\sum_{\left\langle \mu,\nu\right\rangle }\frac{\Lambda_{\mu\nu}}{2}l_{\mu\nu},\label{eq:vertex_model}
\end{equation}
where $N$ is the total number of cells, $A_{i}$ is the area of the
cell $i$, while $A_{i}^{0}$ is its reference area. $K_{i}$
is the area modulus, i.e. a constant with units of energy per area
squared measuring how hard it is to change the cell's area. $P_{i}$ is the cell perimeter and $\Gamma_{i}$ (with units of
energy per length squared) is the perimeter modulus that determines
how hard it is to change perimeter $P_{i}$. $l_{\mu\nu}$ is the
length of the junction between vertices $\mu$ and $\nu$ and $\Lambda_{\mu\nu}$
is the tension of that junction (with units of energy per length).
$\left\langle \mu,\nu\right\rangle$ in the last term denotes the sum is over all pairs of vertices that share a junction.
 Note that the model allows for different cells to have different
 area and perimeter moduli as well as reference areas, allowing for modelling of tissues
containing different cell types. 

Some authors write the perimeter
term as $\sum_{i}\tilde{\Gamma}_{i}/2\left(P_{i}-P_{i}^{0}\right)^{2}$,
where $P_{i}^{0}$ is a reference perimeter, and omit the last term
in Eq.~(\ref{eq:vertex_model}) or completely omit the $P^{2}$ term.\cite{honda1980much,spencer2016vertex}
Under the assumption that the values of $\Lambda_{\mu\nu}$ for all
junctions of the cell $i$ are the same, i.e., $\Lambda_{\mu,\nu}\equiv\Lambda_{i}$,
the last term in Eq.~(\ref{eq:vertex_model}) becomes $\Lambda_{i}\sum_{\left\langle \mu,\nu\right\rangle }l_{\mu,\nu}=\Lambda_{i}P_{i}$.
Therefore, if we identify $\Lambda=-\tilde{\Gamma}P_{i}^{0}$, it
immediately becomes clear that the descriptions in Eq.~(\ref{eq:vertex_model})
and the model with the preferred perimeter are identical to each other.
Note that this is true up to the constant term $1/2\tilde{\Gamma}_{i}\left(P_{i}^{0}\right)^{2}$,
which is unimportant as it only shifts the overall energy and does
not contribute to the force on the cell (see below). The description
in Eq.~(\ref{eq:vertex_model}) is slightly more general
as it allows for the junctions to have different properties depending
on the types of cells that are in contact. Retaining the $P_{i}^{2}$
term is also advisable in order to prevent the model from becoming
unstable if the area modulus is too small. 

It is straightforward to express cell area and cell perimeter in terms
of vertex coordinates. Therefore, vertex positions together with their
connectivities uniquely determine the energy of the epithelial sheet,
hence the name Vertex model. The main assumption of the VM is that
the tissue will always be in a configuration which minimises the total
energy in Eq.~(\ref{eq:vertex_model}). Determining the minimum energy
configuration is a non-trivial multidimensional optimisation problem
and, with the exception of a few very simple cases, it can only be solved
numerically. A basic implementation of the VM therefore needs to use
advanced multidimensional numerical minimisation algorithms to determine
the positions of vertices that minimise Eq.~(\ref{eq:vertex_model})
for a given set of parameters $K_{i}$, $\Gamma_{i}$ and $\Lambda_{\mu\nu}$.
Most implementations,\cite{fletcher2013implementing,spencer2016vertex,Curran078204}
also introduce topology (connectivity) changing moves to model events
such as cell rearrangements.

There have been several attempts to introduce dynamics into the
VM,\cite{honda1980much,weliky1990mechanical,fletcher2013implementing}
including a recent study that introduced stochasticity into the junction
tension.\cite{Curran078204} The idea behind such approaches it to
write equations of motion for each vertex as 
\begin{equation}
\gamma\frac{d\mathbf{r}_{\mu}}{dt}=\mathbf{F}_{\mu},\label{eq:eqation_of_motion_vertex}
\end{equation}
where $\gamma$ is a friction coefficient and $\mathbf{r}_{\mu}$ is
the position vector of vertex $\mu$. $\mathbf{F}_{\mu}$ is the total
force on vertex $\mu$ computed as the negative gradient of Eq.~(\ref{eq:vertex_model})
with respect to $\mathbf{r}_{\mu}$, i.e., $\mathbf{F}_{\mu}=-\nabla_{\mathbf{r}_{\mu}}E_{VM}$.
We note that the exact meaning of friction in confluent epithelial
tissues is the subject of an ongoing debate that is beyond the scope
of this study. Here, as in the case of most models to date, we
assume that all effects of friction (i.e., between neighbouring cells
as well as between cells and the substrate and the extracellular matrix)
can be modelled by a single constant. While this may appear to be a major
simplification, as we will show below, the model is capable of capturing many key features of real epithelial tissues.  Eq.~(\ref{eq:eqation_of_motion_vertex}) is a first order equation
since the mass terms have been omitted. This so-called overdamped
limit is very common in biological systems, since the inertial effects
are typically several orders of magnitude smaller than the effects
arising from the cell-cell interactions or random fluctuations produced
by the environment. Note that the force on vertex $\mu$ depends on
the position of its neighbouring vertices, resulting in a set of coupled
non-linear ordinary differential equations. In most cases those equations
can only be solved numerically.

While the introduction of dynamics alleviates some of the problems
related to the quasi-static nature of the VM, one still has to implement topology changing moves if
the model is to be applicable to describing cell intercalation events.
This can lead to unphysical back and forth flips of the same junction
and has only recently been analysed in full detail.\cite{spencer2016vertex}

\subsection{Active Vertex Model}

It is important to note that the VM in its original form is a \emph{quasi-static}
model. In other words, it assumes that at every instant
in time, the tissue is in a state of mechanical equilibrium. This is a strong assumption,
which is in line with many biological systems, especially
in the case of embryos where cells actively grow, divide and rearrange.
As a matter of fact, biological systems are among the most common
examples of systems out of equilibrium. Therefore, while it is able
to capture many of the mechanical properties of the tissue, the VM
is unable to fully describe the effects that are inherently related
of being out of equilibrium. Many such effects are believed to be
behind the collective migration patterns observed in recent experiments.
In addition, in many dynamical implementations of the VM the effects
of both thermal and non-thermal random fluctuations originating in
complex intercellular processes and interactions with the environment
are either completely omitted or not very clear. While for a system
out of equilibrium the fluctuation-dissipation theorem\cite{chaikin2000principles}
does not hold, and the relation between random fluctuations and friction
is not simple, it is even more important to note that fluctuations can have non-trivial
effects on the collective motion patterns.\cite{Curran078204}

Here we take an alternative approach and build a description similar
to the recently introduced SPV model.\cite{bi2016motility} The idea
behind the SPV is that instead of treating vertices as the degrees
of freedom, one tracks positions of cell centres. Forces on cell centres
are, however, computed from the energy of the VM, Eq.~(\ref{eq:vertex_model}).
The core assumption of the model is that the tissue confirmations correspond
to the Voronoi tessellations of the plane with cell centres acting
as Voronoi seeds. We recall that a Voronoi tessellation is a polygonal
tiling of the plane based on distances to a set of points, called
\emph{seeds}. For each seed point there is a corresponding polygon
consisting of all points closer to that seed point than to any other.
This imposes some restrictions onto possible tissue confirmations,
i.e., not all convex polygonal tessellations of the plane are necessarily
Voronoi, but it has recently been argued that Voronoi tessellations
can predict the diverse cell shape distributions of many tissues.\cite{sanchez2016fundamental}
Furthermore, the exact details of the tessellation are not expected
to play a significant role in the large scale behaviour of the tissue,
which this model aims to describe. 

In the original implementation of Bi, \emph{et al.\cite{bi2016motility}}
the Voronoi tessellation is computed at every time step. The vertices
of the tessellation are then used to evaluate forces at all cell
centres, that are, in turn, moved in accordance to those forces and
the entire process is repeated. While conceptually clear, this procedure
is numerically expensive as it requires computation of the entire
Voronoi diagram at each time step. This limits the accessible system
size to several hundred cells. 

\begin{figure*}
\begin{centering}
\includegraphics[width=0.95\textwidth]{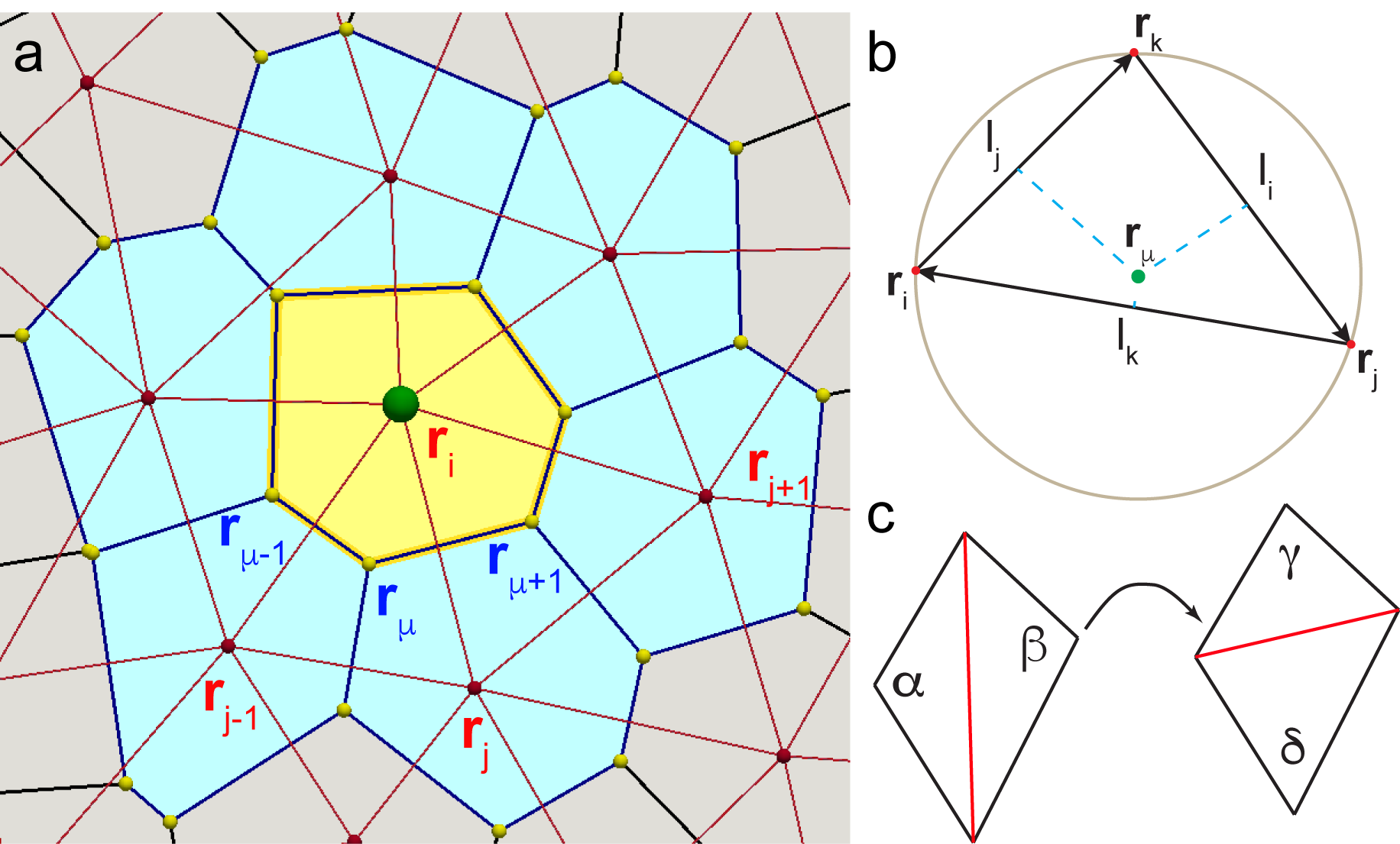} 
\par\end{centering}
\caption{a) Two coordinate representations dual to each other. In the AVM we
track particles that correspond to cell centres (red/green dots).
Particles form a Delaunay triangulation (red lines) and their positions
are labeled with vectors carrying Latin indices. The dual of the Delaunay triangulation is a Voronoi tessellation, with each Voronoi
cell (a shaded polygon) representing an actual cell. Cell edges are
marked by blue lines. Positions of the vertices (yellow dots) of
the dual mesh are denoted by Greek indices. b) Position of the circumcenter
$\mathbf{r}_{\mu}$ of the triangle with corners $\mathbf{r}_{i}$,
$\mathbf{r}_{j}$ and $\mathbf{r}_{k}$. c) The edge flip move is
at the core of the equiangulation procedure. If the sum of the angles
opposite to the red edge exceeds $\alpha+\beta>180^{\circ}$ the edge is
``flipped''. As the result, the sum of the angles opposite to the
new edge is less than $180^{\circ}$, i.e., $\gamma+\delta<180^{\circ}$.
Note that this edge flip is local and only affects one Voronoi edge
(i.e., cell junction). Therefore, edge flips can only affect the local
connectivity of four cells (see also, Fig.~\ref{fig:T1_lapse}). \label{fig:model} }
\end{figure*}

Here, we instead propose an alternative approach based on the Delaunay
triangulation. The Delaunay triangulation for a set of
points $P$ in the plane is a triangular tiling, $DT\left(P\right)$,
of the plane with the property that there are no points of $P$ inside
the circumcircle of any of the triangles in $DT\left(P\right)$.\cite{delaunay1934sphere}
A property of a Delaunay triangulation that is key for this work is
that it is possible to construct a so-called dual Voronoi tessellation by
connecting circumcenters of its triangles. This establishes a mathematical
duality between Delaunay and Voronoi descriptions. This duality is
exact and quantities, such as the force, expressed on the Voronoi
tiling have an exact map onto quantities expressed on its dual Delaunay
triangulation. Although being non-linear (see Sec. \ref{subsec:force_cell}),
this map is relatively simple, and therefore fast to compute. An important
property of the Voronoi-Delaunay duality is that continuous deformations
of one map into continuous deformations of the other. In other words,
smooth motion of a cell's centre will correspond to a smooth change
in that cell's shape. This is crucial to ensuring that during the dynamical
evolution the cell connectivity changes continuously, a feature that
is essential for accurately modelling T1 transitions. The main
advantage of working with the Delaunay description is that while the Voronoi
tessellation has to be recomputed each time cell centres move, it
is possible to retain the Delaunay character of a triangulation via
local edge flip moves (Fig.~\ref{fig:model}c), which drastically
increases the efficiency of the Delaunay based approach and enables
us to simulate systems containing tens of thousands of cells.

Before we introduce the Active Vertex Model (AVM), we pause to make
a comment about the notation. In the following, we will always use
Latin letters to denote cells, i.e. positions of their centres, and
Greek letters to denote vertices of the dual Voronoi tessellation, i.e.,
meeting points of three or more cells. Therefore, vertices of the
VM will always carry Greek indices (Fig.~\ref{fig:model}a).

\subsubsection{Force on the cell centre}

\label{subsec:force_cell}In the AVM, the area $A_{i}$
in Eq.~(\ref{eq:vertex_model}) corresponding to the cell (particle)
$i$ is the area of the polygon (face) of the Voronoi polygon, $\Omega_{i}$,
associated with $i$. It is given as 
\begin{equation}
A_{i}=\frac{1}{2}\sum_{\mu\in\Omega_{i}}\left(\mathbf{r}_{\mu}\times\mathbf{r}_{\mu+1}\right)\cdot\mathbf{N}_{i},\label{eq:area_particle}
\end{equation}
where $\mathbf{r}_{\mu}$ is the position of vertex $\mu$ and $\mathbf{N}_{i}\equiv\mathbf{e}_{z}$
is a unit-length vector perpendicular to the surface of the polygon
 - which does not depend on the position of the vertices in the plane. The sum is over
all vertices of the Voronoi cell and we close the loop with $\mu+1\equiv1$ for
$\mu=N_{\Omega_{i}}$, where $N_{\Omega_{i}}$ is the total number
of vertices in the cell $\Omega_{i}$. This expression is just a discrete
version of Green's formula. Similarly, the cell perimeter is 
\begin{equation}
P_{i}=\sum_{\mu\in\Omega_{i}}\left|\mathbf{r}_{\mu+1}-\mathbf{r}_{\mu}\right|,\label{eq:perim_particle}
\end{equation}
with the same rules for the next neighbour of the ``last'' vertex.
$\left|\cdot\right|$ represents the Euclidean length of the vector.

In order to make a connection between cell centres and indices of
the Voronoi tessellation we recall that for a given triangle in the
Delaunay triangulation, the position of its dual Voronoi vertex coincides
with the centre of the circumscribed circle. The position of the circumcenter
is given by (see Fig.~\ref{fig:model}b) 
\begin{equation}
\mathbf{r}_{\mu}=\frac{\lambda_{1}}{\Lambda}\mathbf{r}_{i}+\frac{\lambda_{2}}{\Lambda}\mathbf{r}_{j}+\frac{\lambda_{3}}{\Lambda}\mathbf{r}_{k},\label{eq:r_mu_baricentric}
\end{equation}
where $\mathbf{r}_{i}$, $\mathbf{r}_{j}$ and $\mathbf{r}_{k}$ are
position vectors of the corners of the triangle and $\lambda_{1}$,
$\lambda_{2}$ and $\lambda_{3}$ are the barycentric coordinates
(see, Eq.~(\ref{eq:circum_lambda}) in Appendix I), with $\Lambda=\lambda_{1}+\lambda_{2}+\lambda_{3}$.

Armed with the mapping in Eq.~(\ref{eq:r_mu_baricentric}), we can
proceed to compute the forces acting on each cell centre based on
the VM energy functional in Eq.~(\ref{eq:vertex_model}). This is
done directly by computing the negative gradient of the energy. Therefore,
the force on the centre of cell $i$ is computed as 
\begin{equation}
\mathbf{F}_{i}=-\nabla_{\mathbf{r}_{i}}E_{VM}.\label{eq:general_force-1}
\end{equation}
The expression for the gradient is made somewhat complicated by the
fact that the derivative is taken with respect to the position of
the cell centre, while the energy of the VM is naturally written in
terms of the positions of the dual vertices. After a lengthy but straightforward
calculation (see Appendix I) we obtain: 
\begin{align}
\mathbf{F}_{i} & =-\sum_{k=1}^{N}\frac{K_{k}}{2}\left(A_{k}-A_{k}^{0}\right)\sum_{\nu\in\Omega_{k}}\left[\mathbf{r}_{\nu+1,\nu-1}\times\mathbf{N}_{k}\right]^{T}\left[\frac{\partial\mathbf{r}_{\nu}}{\partial\mathbf{r}_{i}}\right]\nonumber \\
 & -\sum_{k=1}^{N}\Gamma_{k}P_{k}\sum_{\nu\in\Omega_{k}}\left(\hat{\mathbf{r}}_{\nu,\nu-1}-\hat{\mathbf{r}}_{\nu+1,\nu}\right)^{T}\left[\frac{\partial\mathbf{r}_{\nu}}{\partial\mathbf{r}_{i}}\right]\nonumber \\
 & -\sum_{k=1}^{N}\sum_{\nu\in\Omega_{k}}\left[\Lambda_{\nu-1,\nu}\hat{\mathbf{r}}_{\nu,\nu-1}-\Lambda_{\nu,\nu+1}\hat{\mathbf{r}}_{\nu+1,\nu}\right]^{T}\left[\frac{\partial\mathbf{r}_{\nu}}{\partial\mathbf{r}_{i}}\right].\label{eq:force_expression}
\end{align}
In the last expression, $\left[\frac{\partial\mathbf{r}_{\nu}}{\partial\mathbf{r}_{i}}\right]$
is the $3\times3$ Jacobian matrix (see Appendix I) connecting coordinates
of cell centres with coordinates of the dual Voronoi tessellation and $\left[\cdot\right]^{T}\left[\cdot\right]$
denotes a row-matrix product producing a $3\times1$ column vector.
Note that this product does not commute, i.e., the order in which
terms appear in the expression above is important. $N$ is the total
number of cells in the system. 

We note that most terms in the $k$ sum in Eq.~(\ref{eq:force_expression})
are equal to zero, since each vertex coordinate $\mathbf{r}_{\nu}$ depends only the cell centres $\mathbf{r}_k$ associated to its Delaunay triangle (see Fig.~\ref{fig:model}b). In other words,
 we only need to consider cell $i$
and its immediate neighbours. For clarity, we outline the algorithm
for computing the area term and note that perimeter and junction terms
can be treated in a similar fashion. 
\begin{enumerate}
\item For particle $i$ compute $\frac{K_{i}}{2}\left(A_{i}-A_{i}^{0}\right)$
and multiply it by the sum $\sum_{\nu\in\Omega_{i}}\left[\left(\mathbf{r}_{\nu+1}-\mathbf{r}_{\nu-1}\right)\times\mathbf{N}_{i}\right]^{T}\left[\frac{\partial\mathbf{r}_{\nu}}{\partial\mathbf{r}_{i}}\right]$.
This sum is over all vertices (corners) $\nu$ of cell $i$. 
\item For all immediate neighbours $j$ of cell $i$ compute $\frac{K_{j}}{2}\left(A_{j}-A_{j}^{0}\right)$
and multiply it by the sum 
\[
\sum_{\nu\in\Omega_{i}\cap\Omega_{j}}\left[\left(\mathbf{r}_{\nu+1}-\mathbf{r}_{\nu-1}\right)\times\mathbf{N}_{j}\right]^{T}\left[\frac{\partial\mathbf{r}_{\nu}}{\partial\mathbf{r}_{i}}\right].
\]
Note that $\nu\in\Omega_{i}\cap\Omega_{j}$ ensures that vertices
$\nu$ surrounding $j$ are taken into account only if they are affected
by (i.e., also belong to) the cell $i$.  
\end{enumerate}
From the previous discussion it is clear that the expression for the
force is local, i.e., computing the force does not require including
cell centre positions beyond the immediate neighbourhood of a given
cell. This is extremely beneficial from a computational point of view
as one can readily utilise standard force cutoff techniques, such
as cell and neighbour lists\cite{allen1989computer} in order to speed
up force computations. It is also evident that the force in Eq.~(\ref{eq:force_expression})
is not pairwise, i.e., it cannot be written as a sum of forces acting
on a pair of cells, or in mathematical terms, $\mathbf{F}_{i}\neq\sum_{j}\mathbf{F}_{ij}$.
This is not surprising since the position of the Voronoi vertices
depends on the position of three cell centres.

\subsubsection{Cell alignment}

\label{subsec:alignment}The AVM also includes the\emph{ cell polarity\cite{asnacios2012mechanics}}
modelled as a unit length vector, $\mathbf{n}_{i}$, laying in the
$xy$ plane. This vector determines the direction of the cell's motion
and division and should not be confused with apical-basal polarity,
which is not included in this model. In the simplest case, the vector
$\mathbf{n}_{i}$ does not take any preferred direction, but instead
randomly fluctuates under the influence of uncorrelated random noise
originating from the intercellular processes and the environment. This
simple model can be augmented to include several different models
for cell-cell alignment. For example, a cell can be given a tendency
to align its polarity to its immediate neighbours by minimising the
alignment energy, 
\begin{equation}
E_{polar\,align}=-J_p\sum_{j\,\mathrm{n.n}\,i}\mathbf{n}_{i}\cdot\mathbf{n}_{j},\label{eq:xy_aligment}
\end{equation}
where $J_p>0$ is the alignment strength and the sum is over all nearest
neighbours of $i$. This model is similar to the Vicsek model, which
is used in studies of flocking, e.g. of swarms of birds.\cite{chate2008modeling}
The molecular details of polar alignment are not yet understood
in detail. It as been argued that in some tissue systems this involves
components of the planar cell polarity signalling cascade,\cite{zallen2007planar,wallingford2012planar}
but even here their precise coordination and involvement are as yet
not resolved.

Alternative alignment mechanisms consider only internal processes
in the cell and do not directly depend on the polarity of the neighbouring
cells. One such mechanism introduced in physics of active matter\cite{szabo2006phase,henkes2011active}
assumes that cell polarisation vector aligns with the migration direction
of the cell, i.e., 
\begin{equation}
E_{vel\,align}=-J_{v}\:\mathbf{n}_{i}\cdot\mathbf{\hat{v}}_{i},\label{eq:velocity_align}
\end{equation}
where $\mathbf{\hat{v}}_{i}$ is the normalised cell velocity vector
and $J_{v}>0$ is the alignment strength.

Finally, $\mathbf{n}_{i}$ can also align to the direction of the
eigenvector of the cell shape tensor, i.e. the cell seeks align direction
of $\mathbf{n}_{i}$ along its long axis. This is achieved by minimising
the value of the energy 
\begin{equation}
E_{shape\,align}=-J_{s}\mathbf{n}_{i}\cdot\mathbf{p}_{i},\label{eq:shape_align}
\end{equation}
where $J_{s}>0$ is the alignment strength and $\mathbf{p}_{i}$ is
the eigenvector of the matrix\cite{aubouy2003texture} 
\begin{equation}
\hat{P}_{i}=\frac{1}{N_{\Omega_{i}}}\sum_{\mu\in\Omega_{i}}\mathbf{r}_{\mu}\otimes\mathbf{r}_{\mu+1}=\frac{1}{N_{\Omega_{i}}}\sum_{\mu\in\Omega_{i}}\left(\begin{array}{cc}
\left(x_{\mu}-x_{\mu+1}\right)^{2} & \left(x_{\mu}-x_{\mu+1}\right)\left(y_{\mu}-y_{\mu+1}\right)\\
\left(x_{\mu}-x_{\mu+1}\right)\left(y_{\mu}-y_{\mu+1}\right) & \left(y_{\mu}-y_{\mu+1}\right)^{2}
\end{array}\right),\label{eq:shape_tensor}
\end{equation}
corresponding to its largest eigenvalue. This mechanism is one possible
pathway to obtain \emph{plithotaxis}.

Each of these alignment mechanisms causes a torque on $\mathbf{n}_{i}$,
given by 
\begin{equation}
\boldsymbol{\tau}_{i}=-\mathbf{n}_{i}\times\nabla_{\mathbf{n}_{i}}E_{align}.\label{eq:torque}
\end{equation}
In the \emph{SAMoS} implementation of the AVM, one can select which of these alignment mechanisms are to be included in an actual simulation.

\subsubsection{Equations of motion}

We proceed to write equations of motion for the position and orientation
of cell $i$ in the AVM. As discussed in Sec. \ref{subsec:overview},
at the cellular scale inertial effects are negligible. In this overdamped
limit, the equation of motion for the position of the cell centre is just force balance between friction and driving forces. If we assume
that the friction of the cell with its surrounding and the substrate
is isotropic and can be modelled by the single friction coefficient
$\gamma$, the equation of motion for the position of cell centre $i$ becomes 
\begin{equation}
\gamma\frac{d\mathbf{r}_{i}}{dt}=f_{\text{a}}\mathbf{n}_{i}+\mathbf{F}_{i}+\boldsymbol{\nu}_{i}\left(t\right),\label{eq:motion_postition}
\end{equation}
where $\mathbf{F}_{i}$ is the sum of all forces acting on the cell,
i.e., a sum of the VM model forces given in Eq.~(\ref{eq:force_expression})
and the soft-core repulsion defined in Eq.~(\ref{eq:v_soft}) in Appendix
I. $\boldsymbol{\nu}_{i}\text{\ensuremath{\left(t\right)}}$ is an
uncorrelated stochastic force that models the effects of random fluctuations.
These fluctuations originate from the intracellular processes and interactions
with the environment. In its simplest possible form, one can assume
that $\boldsymbol{\nu}_{i}\text{\ensuremath{\left(t\right)}}$ has
no time correlations, i.e., each cell experiences a random ``kick''
at a given instant of time whose magnitude and direction do not
depend on any past ``kicks'' on that cell. Formally, we write $\left\langle \nu_{i,\alpha}\left(t\right)\right\rangle =0$
and $\left\langle \nu_{i,\alpha}\left(t\right)\nu_{j,\beta}\left(t'\right)\right\rangle =2D\delta_{ij}\delta_{\alpha\beta}\delta\left(t-t'\right)$,
where $D$ measures the average magnitude of the stochastic force
and is interpreted as an effective translational diffusion coefficient,
$\alpha,\beta\in\left\{ x,y\right\} $ and $\left\langle \dots\right\rangle $
denotes the statistical average. Finally, the term $f_{\text{a}}\mathbf{n}_{i}$
has its origins in the physics of active matter systems.\cite{marchetti2013hydrodynamics}
It models \emph{activity}, i.e., internal cellular processes that
drive it to move in the direction of its polarity. The active force
strength $f_{\text{a}}$ controls the magnitude of this activity.
While the biological meaning of $f_{\text{a}}$ may appear unclear,
it quantifies a cell's ability to move on its own due to the complex molecular
machinery within it. In many models of individual cells crawling
on a substrate with a prominent lamellipodium, the resultant active
velocity $v_{0}=f_{\text{a}}/\gamma$ is due either actin tread-milling,\cite{du2005force}
differential friction or differential contractility. In an epithelium,
on the other hand, $f_{\text{a}}$ should be understood as an effective
parameter that models the non-balanced active force due to junction
contractions and internal cell contractility.\cite{biname2010makes}
Although the molecular origin of $f_{a}$ is at present not fully
understood, even this simple description of the activity combined
with the cell polarity alignment models discussed in the previous subsection
leads to collective behaviour that resembles behaviour observed in
experiments. 

We then write equations for motion for the polarity vector. If we
define the angle of vector $\mathbf{n}_{i}$ with the $x-$axis as
$\theta_{i}$, then $\mathbf{n}_{i}=\left(\cos\theta_{i},\sin\theta_{i}\right)$
and we have 
\begin{equation}
\gamma_{r}\frac{d\theta_{i}}{dt}=\boldsymbol{\tau}_{i}\cdot\mathbf{N}_{i}+\nu_{i}^{r}\left(t\right),\label{eq:motion_direction}
\end{equation}
where $\boldsymbol{\tau}_{i}$ is the torque acting on $\mathbf{n}_{i}$,
given in Eq.~(\ref{eq:torque}), $\mathbf{N}_{i}$ is the local normal
to the cell surface (i.e., simply the unit length vector in the $z$-direction),
$\gamma_{r}$ is the orientational friction, and $\nu_{i}^{r}\left(t\right)$
is responsible for introducing an orientational randomness. Akin to
the stochastic force $\boldsymbol{\nu}_{i}\text{\ensuremath{\left(t\right)}}$
in Eq.~(\ref{eq:motion_postition}), $\nu_{i}^{r}\left(t\right)$
has properties $\left\langle \nu_{i}^{r}\left(t\right)\right\rangle =0$
and $\left\langle \nu_{i}^{r}\left(t\right)\nu_{j}^{r}\left(t'\right)\right\rangle =2D_{r}\delta_{ij}\delta\left(t-t'\right)$,
where $D_{r}$ is an rotational diffusion constant. The related time
scale $\tau_{r}=\gamma_{r}/2D$ measures the time necessary for the
cell polarisation direction to decorrelate due to fluctuations produced by cellular
processes and the environment. The unit-length vector $\mathbf{n}_{i}$
then evolves according to $\frac{d\mathbf{n}_{i}}{dt}=\frac{d\theta_{i}}{dt}\mathbf{N}_{i}\times\mathbf{n}_{i}$.\cite{sknepnek2015active}

Eqs. (\ref{eq:motion_postition}) and (\ref{eq:motion_direction})
describe the cell dynamics in the AVM. These equations are solved
numerically using several standard time discretisation schemes. 

\subsubsection{Cell growth, division and death}

\label{subsec:grow-div-death}Another contribution to the activity
comes from cell growth, division and death or extrusion from the cell
sheet, present in many epithelial tissues. We will refer to these
three processes as \emph{population control}. Cell growth is modelled
as a constant-rate increase of the native area, i.e., 
\begin{equation}
A_{0}\left(t+\delta t\right)=\left(1+\eta\delta t\right)A_{0}\left(t\right),\label{eq:growth}
\end{equation}
where $\eta$ is the growth rate per unit time and $\delta t$ is
the simulation time step. Cell death is modelled by tracking the cell's
``age''. Age is is increased at every time step by $\delta t$.
Once the age reaches a critical value, the cell is removed from the
system. We note that onset of the cell's death (or alternative its extrusion from the sheet) may also depend on the
stresses or forces exerted on the cell. In the current implementation
of the AVM we do not include such effects. Adding such effects to the
model would be straightforward. 

Finally, cell division is modelled based on the ideas of Bell and
Anderson.\cite{bell1967cell} The cell can only divide once its area
is larger then some critical area $A_{c}$, in which case it divides
with probability proportional to $A-A_{c}$, i.e., 
\begin{equation}
p_{div}=\begin{cases}
\chi\left(A-A_{c}\right) & \,\,\mathrm{for}\,\,A>A_{c}\\
0 & \,\,\mathrm{otherwise}
\end{cases},\label{eq:div_prob}
\end{equation}
where $\chi$ is a constant with units of inverse area times time.
Upon division the two daughter cells are placed along the direction
of the vector $\mathbf{n}_{i}$ and their ages are reset to zero.
The Delaunay triangulation is rebuilt after every division or death
event.

The cell cycle is clearly a far more complex phenomenon than what is captured
by this simple model. The simplest possible extension would be to
build upon the ideas of Smith and Martin,\cite{smith1973cells} or
use many other more sophisticated models available in the literature.
Adding such extensions to the AVM would be straightforward and will
be included in later versions of the model. 

\subsubsection{Maintaining a Delaunay triangulation/Voronoi tessellation}

\label{subsec:equiangulation}In an actual simulation, we start either
from carefully constructed initial positions of cell centres or choose
cell positions from a particular experimental system. Those positions
are then used to build the initial Delaunay triangulation and its
corresponding dual Voronoi tessellation. However, as the cell centres
move, there is no guarantee that the Delaunay character of the triangulation
is preserved. For the model to be able to properly capture cell dynamics
we, however, need to ensure that the triangulation is indeed Delaunay
at each time step. Building a new triangulation in every time step
would be computationally costly. Instead, we apply the so-called equiangulation
procedure:\cite{brakke1992surface} For every edge in the triangulation
we compute the sum of the angles opposite to it. If the sum is larger
than $180^{\circ}$ we ``flip'' the edge (see Fig.~\ref{fig:model}c).
The procedure is repeated until there are no more edges left to flip.
One can show\cite{brakke1992surface} that this procedure always converges
and leads to a Delaunay triangulation. While equiangulation is
not the most efficient way to build a Delaunay triangulation ``from
scratch'', if one starts with a triangulation that is nearly Delaunay,
in practice only a handful of flip moves are required to recover the
Delaunay triangulation. Given that cell centres move continuously,
in the AVM the equiangulation procedure significantly increases the
efficiency of maintaining the Delaunay triangulation throughout the
simulation.

It is important to note that edge flips are local and flipping one
edge in the Delaunay triangulation only affects one edge of the dual
Voronoi tessellation. This means that a single edge flip can only
affect junctions between four cells. The centres of those four cells coincide
with the locations of the four corners of the polygon formed by two
triangles sharing the flipped edge (see, Fig.~\ref{fig:model}c).
This is precisely the mechanism behind the T1 transition discussed
in Sec. \ref{subsec:T1_transitions} below. No other cell contacts
are affected by the flip.

\subsubsection{Handling boundaries}

The implementation of the SPV by Bi, \emph{et al}.\cite{bi2016motility}
assumes periodic boundary conditions. This assumption is reasonable
if one studies a relatively small region of a much larger epithelial
tissue. Many experiments, especially those of cell migration on elastic
substrates, however, are performed with a relatively small number
of cells where the effects of boundaries cannot be neglected or are
even the main focus of the study. Therefore, in the AVM we include
an open, flexible boundary.

\begin{wrapfigure}{r}{0.5\textwidth}
\begin{centering}
\includegraphics[width=0.5\textwidth]{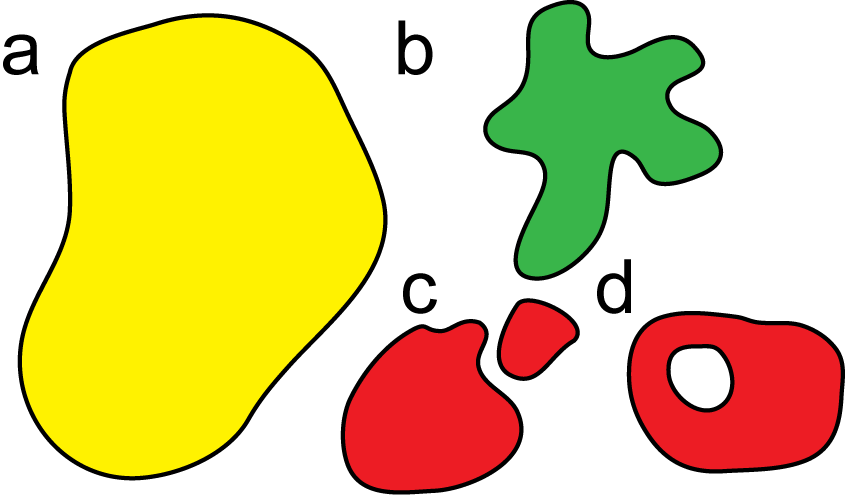} 
\par\end{centering}
\caption{Schematic representation of the allowed and disallowed changes of
the boundary in the AVM. An initial configuration with the topology
of a disk (a) is allowed to develop pronounced fingers (b). However,
it is not possible to split into two domains (c) or develop a hole
(d), which would both lead to the introduction of a new boundary lines,
and therefore lead to changes in the topology. \label{fig:boundary_topology}}
\end{wrapfigure}

In order to avoid very costly checks of topology changes, we assume
that the topology (connectivity) of the boundary is maintained throughout
the entire simulation. This does not mean that the boundary is fixed.
It can grow, shrink and change its shape, but it cannot change its
topology, e.g. it is not possible to transition from a disk to an
annulus. Examples of allowed and disallowed changes of the boundary
are shown in Fig.~\ref{fig:boundary_topology}. While fixing the boundary
topology may appear restrictive, we will see below that in practice
a model with fixed topology of the boundary allows for detailed studies of a broad range of problems directly applicable to many current experiments.

In the AVM, the degrees of freedom are cell centres, represented as
particles. These particles serve as sites of the Delaunay triangulation.
In order to handle boundaries we introduce a special type of \emph{boundary}
particle and in addition, we also specify the connectivity between the
boundary particles. Boundary particles together with their connectivity
information form a \emph{boundary line}. This line sets the topology
of the boundary, which is preserved throughout the simulation, and
delineates between the tissue and its surrounding. In addition, the
boundary line can have an energy associated to it. Biologically, this
boundary energy corresponds to the complex molecular machinery, such
as actin cables, that is known to affect the behaviour of the free
edge of an epithelial sheet. Specifically, we introduce the boundary
line tension, 
\begin{equation}
E_{lt}=\frac{1}{2}\sum_{\left\langle i,j\right\rangle }\lambda_{i,j}\left(l_{ij}-l_{0}\right)^{2}\label{eq:boundary_line_tension}
\end{equation}
where $\lambda_{i,j}$ is the line tension of the edge connecting
boundary vertices $i$ and $j$, $l_{ij}=\left|\mathbf{r}_{i}-\mathbf{r}_{j}\right|$
is the length of that edge and $l_{0}$ is its native (preferred)
length. In addition, we also introduce the boundary bending stiffness
\begin{equation}
E_{bbend}=\frac{1}{2}\sum_{i}\zeta_{i}\left(\theta_{i}-\pi \right)^{2},\label{eq:boudnary_bending}
\end{equation}
where $\zeta_{i}$ is the stiffness of angle $\theta_{i}$ at the
boundary particle $i$, determined as 
\begin{equation}
\theta_{i}=\arccos\frac{\mathbf{r}_{ji}\cdot\mathbf{r}_{ki}}{\left|\mathbf{r}_{ji}\right|\left|\mathbf{r}_{ki}\right|},\label{eq:boundary_angle}
\end{equation}
where $\mathbf{r}_{j}$ and $\mathbf{r}_{k}$ are the positions of
boundary particles to the left and to the right of particle $\mathbf{r}_{i}$.
For simplicity, we assume that each boundary particle has exactly
two boundary neighbours. This prevents somewhat pathological, ``cross-like''
configurations where two otherwise disjoint domains would hinge on
a single boundary site.

It is important to note that boundary particles do not represent centres
of an actual cell. The  Voronoi polygon dual to a boundary particle in
a Delaunay triangulation extends out to infinity. Consequently, it is not possible
to unambiguously assign quantities such as the associated area or
perimeter to boundary particles. Therefore, only internal particles
correspond to cell centres, while boundary particles should be thought
of as ``ghosts'' that serve to mark the edge of the cell sheet.
These particles, however, experience forces from the interior of the
tissue as well as from the interactions with their neighbours on the
boundary. The boundary line is able to dynamically adjust its
shape and length in this manner. We also allow for boundary particles to be added to or removed from the boundary line. Details of the algorithms used to dynamically update
the boundary are given in Appendix II.

\begin{figure}
\begin{centering}
\includegraphics[width=0.45\textwidth]{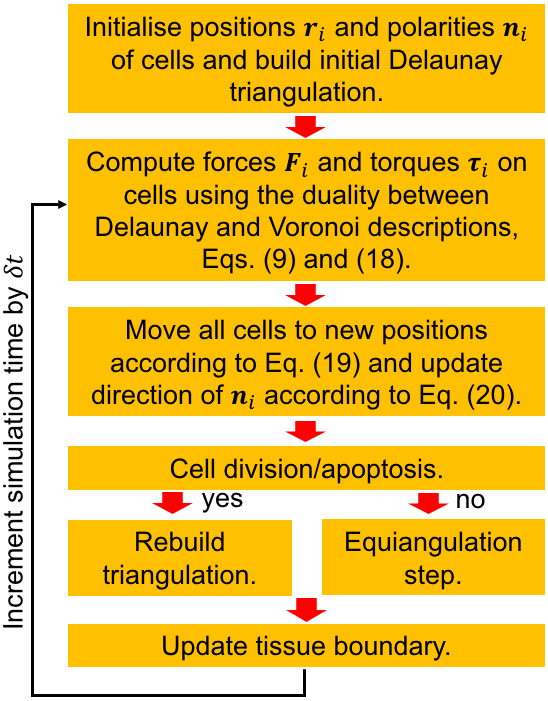}
\par\end{centering}
\caption{Flow chart of the force and torque calculations in each time step
based on Eqs. (\ref{eq:force_expression}) and (\ref{eq:torque}).
The calculation speed is significantly improved by using the fast
equiangulation moves shown in Fig.~\ref{fig:model}c. The Delaunay triangulation
is recomputed only if there are cell division and/or death events,
which do not occur at every time step. \label{fig:flow_chart}}
\end{figure}

This concludes the description of the AVM. In Fig.~\ref{fig:flow_chart}
we present the flow chart of the main steps in computing the time
evolution of cell centres and polarity vectors. Technical details
of the implementation are given in Appendix III and in the on-line
documentation provided with the \emph{SAMoS} code. 

\section{Examples and applications}

\label{sec:examples}

In order to validate the model and compare it with the results of
similar models proposed in the literature, as well as to show its use in modelling actual biological tissues,
we now apply the AVM to several problems relevant to the
mechanics of epithelial tissue layers.

\subsection{T1 transitions}

\label{subsec:T1_transitions}We start by illustrating one of the
key processes observed in epithelial tissues, the T1 transition. As
detailed in Fig.~\ref{fig:model}, in the AVM T1 transitions are handled
through an edge flip in the Delaunay triangulation. An edge flip only
happens when, in the notation of Fig.~\ref{fig:model}c, we have $\alpha+\beta=\delta+\gamma=180^{\circ}$.
Then both triangles are circumscribed by the same circle passing through
its combined four vertices. The location of the T1 transition coincides
with the centre of this circle. Due to the continuous connection between
the position of sites of the Delaunay triangulation and its dual Voronoi
tessellation, we always approach this point smoothly, i.e. a junction
between two cells will smoothly shrink to a point, the T1 transition
will occur, and then it will expand in a new direction. This process
arises naturally in the AVM model, in stark contrast with many currently
available implementations\cite{coburn2016contact} that require a
cut-off criterion on the edge length of a cell before a T1 transition
can occur. It also avoids discontinuous jumps at finite edge length,
bypassing the T1 point altogether, also a feature of a number of models,
notably those based on sequential energy minimisation.\cite{bi2014energy,aegerter2010exploring,farhadifar2007influence}
Next to its high computational efficiency, the ability to smoothly
go through a T1 transition without the need for any additional manipulations
of either the Delaunay triangulation or the Voronoi tessellation is
one of the key advantages of the AVM approach. 

\begin{figure}[h]
\begin{centering}
\includegraphics[width=1\textwidth]{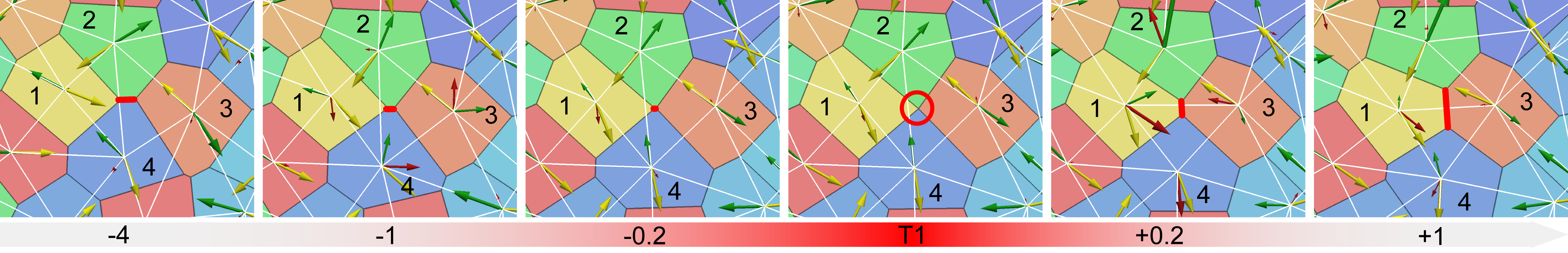} 
\par\end{centering}
\caption{Time lapse of a T1 transition, each label is the time (in units of
$\gamma/K a^{2}$, with $a=1$) with respect to
the T1 event. Initially, cells 2 and 4 are in contact (red line).
Approaching the transition, the connecting line slowly contracts,
until it becomes a point at the transition. Cells 1 and 3 make a new
contact which then rapidly expands. The arrows represent the force
on the cell centre resulting from the Vertex potential (green), which
partially compensates the active self-propulsion force (yellow) to
give the resultant total force (red). The Voronoi tessellation is
outlined in black, and the Delaunay triangulation is in white. \label{fig:T1_lapse}}
\end{figure}

In Fig.~\ref{fig:T1_lapse}, we illustrate a T1 transition in the
bulk, in a region of phase space where the system exhibits liquid-like
behaviour, but with very slow dynamics (see next section). The edge
linking cells 2 and 4 (in red) slowly shrinks to a point, and then
rapidly expands in the opposite direction. This feature points to
dynamics akin to certain models of sheared materials,\cite{hebraud1997yielding}
where the active driving pulls the material over an energy barrier
from one minimum to the next. It is somewhat different from the activated
dynamics which has been proposed for the SPV,\cite{bi2014energy}
which would predict a series of fluctuations through which the barrier
between minima is ultimately crossed.

\subsection{Activity driven fluidisation phase diagram}

\label{subsec:phase_diagram}We now explore different modes of collective behaviour,
i.e., \emph{phases}, of the tissue based on the values of parameters
of the original VM ($K_{i}$, $\Gamma_{i}$ and $\Lambda_{\mu,\nu}$),
and AVM-specific parameters such as the activity $f_{a}$, the orientational correlation time $\tau_r$, and the boundary line tension $\lambda$.
In order to keep the number of independent parameters to a minimum, it is again convenient to rewrite the energy of the
VM, Eq.~(\ref{eq:vertex_model}), in a scaled form.\cite{farhadifar2007influence,bi2016motility}
We first choose $K_{i}=1$ and set $A_{i}^{0}=\pi$ as an area scale. For simplicity, we assume that all perimeter and junction tensions
are the same, i.e., we set $\Gamma_{i}\equiv\Gamma$ and $\Lambda_{\mu,\nu}\equiv\Lambda$
for all $i$, $\mu$ and $\nu$. Then, as discussed below Eq.~(\ref{eq:vertex_model}), we can complete the square on the second and third terms in Eq.~(\ref{eq:vertex_model}) 
and obtain the scaled VM potential
\begin{equation}
E_{VM}=\sum_{i=1}^{N}\frac{1}{2}\left(A_{i}-A_0\right)^{2}+\sum_{i=1}^{N}\frac{\Gamma}{2}\left(P_{i}-P_{0}\right)^{2},\label{eq:vertex_model_scaled}
\end{equation}
where $P_{0}=-\Lambda/\Gamma$ and $A_0=\pi$. The first term in Eq.~(\ref{eq:vertex_model_scaled})
penalises changes in the cell area, while the second term penalises
changes of the perimeter. There is no reason for
the preferred area $A_{0}$ to be generically compatible with the preferred perimeter
$P_{0}$. This sets up a competition between the two terms in Eq.~(\ref{eq:vertex_model_scaled}),
giving a natural scale that is is determined by the relative ratio of $\Gamma P_{0}^{2}$ to
$K A_{0}^{2}$. In other words, if $K A_{0}^{2}>\Gamma P_{0}^{2}$, the cell will try to optimise
its area at the expense of paying a penalty for not having
the most optimal perimeter, and the opposite if $K A_{0}^{2}<\Gamma P_{0}^{2}$. 

Bi, \emph{et al.}\cite{bi2016motility} introduced the dimensionless
\emph{shape factor} $p_{0}=\frac{P_{0}}{\sqrt{A_{0}}}$, which controls the ratio of the cell's
perimeter to its area through the target area $A_0$ and target perimeter $P_0$. The value of $p_{0}$ then determines whether the
area or the perimeter term in Eq.~(\ref{eq:vertex_model_scaled})
wins and effectively sets the preferred shape of each cell:
cells of different shapes have different values of $p_{0}$. For example, regular hexagons, pentagons, squares and triangles corresponds to $p_{0}=3.722$, $p_{0}=3.812$, $p_{0}=4.0$ and $p_{0}=4.559$, respectively. 

Remarkably, one observes\cite{farhadifar2007influence,bi2014energy,bi2016motility}
a transition between a solid-like behaviour of the tissue, where cells do not exchange neighbours,
and liquid-like behaviour, where neighbour exchanges do occur,
at $p_{0}=3.812$, a value that corresponds to a regular pentagon.
At present, the biological significance of this observation is not clear,
but it appears to be a robust feature of many experimental systems.\cite{park2015unjamming}
In order make the comparison between the AVM and the SPV model, we
also adopt $p_{0}$ as a main parameter that controls the preferred
cell shape. 

\begin{figure}[t]
\begin{centering}
\includegraphics[width=1\textwidth]{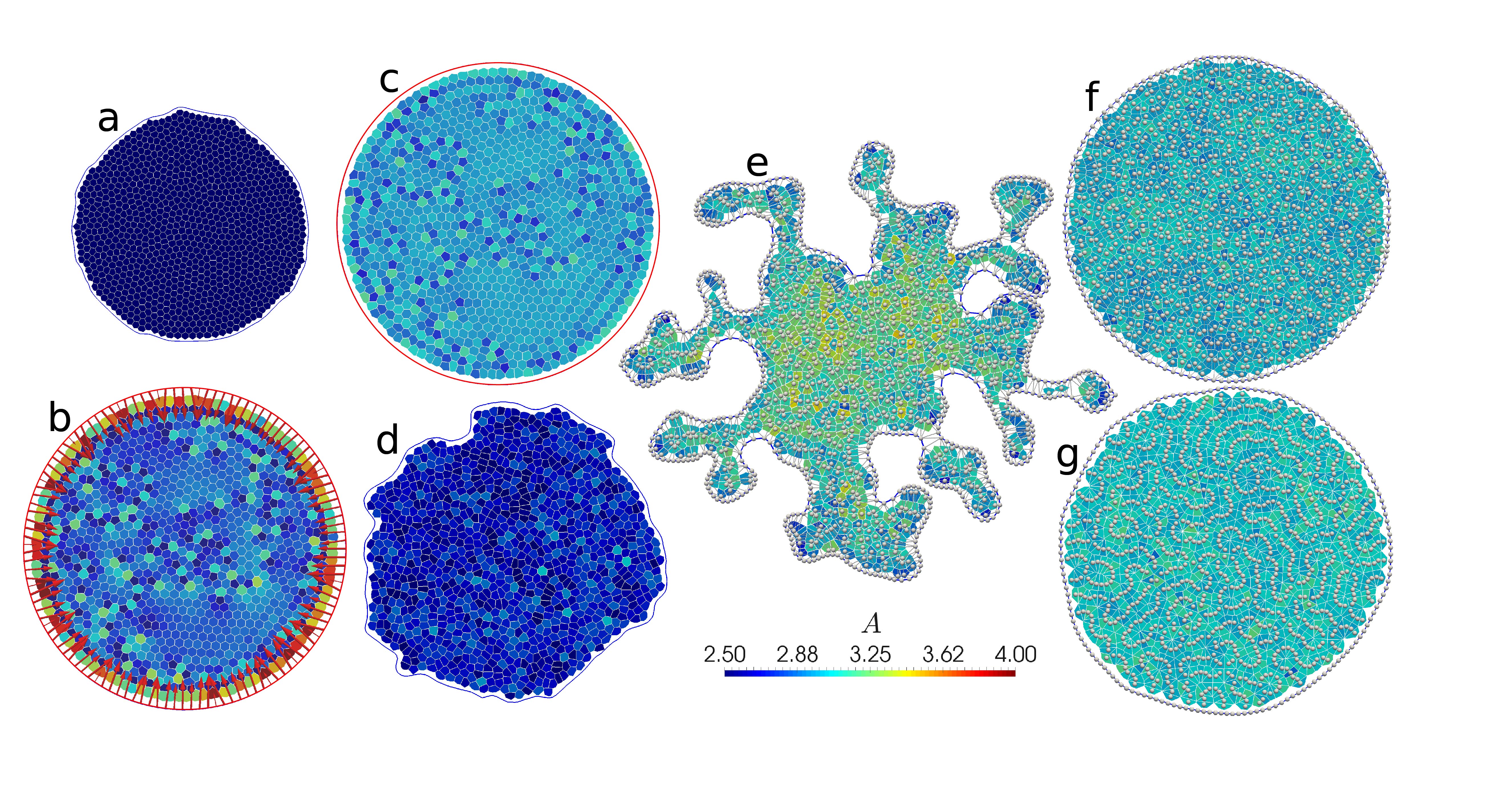} 
\par\end{centering}
\caption{Representative snapshots of states seen with fixed and flexible boundaries in the
AVM. Contact lines between cells are outlined in white, and cells
are coloured according to their area. The line connecting tissue boundary
points is blue for flexible boundaries (panels a, d, e, f
and g), and red for fixed ones (panels b and c). In panels e,
f and g, cell centres are denoted by white spheres, and in panels
e and f, the Delaunay triangulation is also shown as grey lines.
(a) Shrinking cells at $p_{0}=2.48$, $\Gamma=1.0$, $\tau_{r}^{-1}=0.01$
and $f_{\text{a}}=0.1$, no boundary line tension. (b) Same
as (a), but for a fixed boundary. (c) Solid-like (glassy) state at
$p_{0}=3.39$, $\Gamma=1.0$, $\tau_{r}^{-1}=0.1$ and $f_{\text{a}}=0.03$.
(d) Liquid-like state with a fluctuating boundary at $p_{0}=3.39$,
$\Gamma=1.0$, $\tau_{r}^{-1}=0.1$ and $f_{\text{a}}=0.3$,
no boundary line tension (e) Fingering instability at $p_{0}=3.72$,
$\Gamma=0.1$, $\tau_{r}^{-1}=0.01$ and $f_{\text{a}}=0.3$,
boundary line tension $\lambda=0.1$. (f) Fluid state at $p_{0}=3.95$,
$\Gamma=0.1$, $\tau_{r}^{-1}=0.1$ and $f_{\text{a}}=0.1$,
boundary line tension $\lambda=0.3$. (g) Rosette formation at $p_{0}=4.85$,
$\Gamma=0.1$, $\tau_{r}^{-1}=0.1$ and $f_{\text{a}}=0.03$,
boundary line tension $\lambda=0.3$. \label{fig:gallery}}
\end{figure}

In order to initialise the simulation, in each run, we start by placing
soft spheres with slightly polydisperse radii in a circular region. We then use \emph{SAMoS} to minimise
the energy of a soft sphere packing in the presence of a fixed boundary.
This ensures that initially, cells are evenly spaced without being
on a grid. We also fix the packing fraction to $\phi=1$, ensuring that the
average cell area of the initial configuration is $\langle A\rangle=A_{0}$.
The boundary is either fixed (referred to as ``fixed system''),
or allowed to fluctuate freely (``open system''). Fig.~\ref{fig:gallery}
shows a representative set example of the states that we observe. We run the simulation
for either $100,000$ time steps with step size $\delta t=0.01$ in
the unstable region (e.g., Fig.~\ref{fig:gallery}e), or $250,000$
time steps with $\delta t=0.025$ in the solid-like region (e.g.,
Fig.~\ref{fig:gallery}c). For these systems with $N=1000$ cells in the interior, this takes between 10-40 minutes
on a single core of a modern Intel Xeon processor depending on
 the number of rebuilds of the Delaunay triangulation that are necessary (more in the liquid-like phase). 

The unit of time is set by $\gamma/K a^{2}$,
where $a\equiv1$ is the unit of length. We note that Bi, \emph{et
al.} use $a=\sqrt{A_{0}}$ as the unit of length. This is possible
as long as cells are not allowed to grow, i.e., when $A_{0}$ changes
in time. 
The AVM allows for the cells to change their size and therefore we
need to choose a different unit of length. In our case, $a$ is the range of the of soft-core repulsion between cell centres (see, Eq.~(\ref{eq:v_soft}) in Appendix I).

At low values of $p_{0}$, we find a system that prefers to be in
a state with mostly hexagonal cells, unless the active driving $f_{\text{a}}$
is very high. Open systems will shrink at this point so that all cells
are close to their target $P_{0}$, as shown in Fig.~\ref{fig:gallery}a.
Consistent with this, larger values of the perimeter modulus $\Gamma$
lead to stronger shrinking. For fixed systems, this route is blocked,
and instead there is a strong inward tension on the boundaries and
a gradient in local density, as shown in Fig.~\ref{fig:gallery}b. 

\begin{figure}
\begin{centering}
\includegraphics[width=0.95\textwidth]{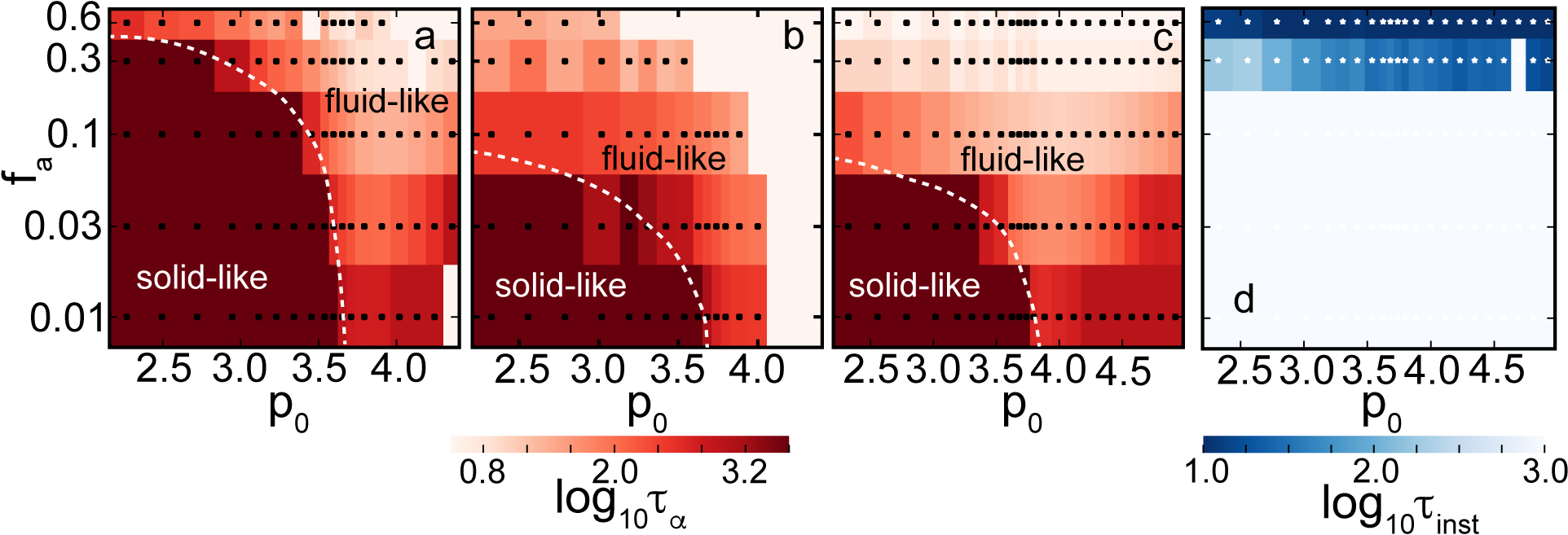} 
\par\end{centering}
\caption{Phase diagrams of the AVM. Panels (a), (b) and (c): $\alpha-$relaxation
time $\tau_{\alpha}$ determined from the self-intermediate scattering
function, Eq.~(\ref{eq:self-intermediate}). These plots indicate
that it is possible to initiate cell intercalation events by changing
values of $p_{0}$ or active driving. High values of $\tau_{\alpha}$
(dark regions) correspond to the solid-like (glassy) phase where T1 events
are suppressed. (a) Fixed system with $\Gamma=1$ and $\tau_{r}^{-1}=0.01$.
(b) Open system with $\Gamma=1$, $\tau_{r}^{-1}=0.01$ and boundary
line tension $\lambda=0$. (c) Open system with $\Gamma=0.1$, $\tau_{r}^{-1}=0.1$
and boundary line tension $\lambda=0.1$. All systems are solid-like
at low $p_{0}$ and low driving $f_{\text{a}}$. The critical
$p_{0}=3.81$ is the same in (a), (b) and (c), but the open system
becomes liquid-like at much lower values of active driving. Lowering
$\Gamma$ also lowers this transition point. The dashed white line represents
a rough boundary between solid-like and fluid-like behaviour. (d)
Characteristic time scale $\tau_{\text{inst}}$ needed to reach the
boundary instability, determined from reaching a threshold boundary
length (see text), for the same parameters as (c). Sufficiently high
driving always leads to an unstable system. \label{fig:phase_diagram}}
\end{figure}

In agreement with the results of Bi, \emph{et al.},\cite{bi2016motility}
we find that at low $p_{0}<3.81$ and low values of driving $f_{\text{a}}$,
cells do not take an organised pattern and do not exchange neighbours.
Recast in the language of solid state physics, the tissue is in an
\emph{amorphous solid} or \emph{glassy} state. In Fig.~\ref{fig:gallery}c
we show such a state for a fixed boundary. In order to characterise the
physical properties of this state, we measure the dynamical time scale
of cell rearrangements through a standard tool of the physics of glassy
systems, the self-intermediate scattering function\cite{berthier2011dynamical}
\begin{equation}
F\left(q,t\right)=\left\langle \exp\left[i\mathbf{q}\cdot\left(\mathbf{r}(t)-\mathbf{r}(0)\right)\right]\right\rangle .\label{eq:self-intermediate}
\end{equation}
$F\left(q,t\right)$ measures the decay of the autocorrelation of cell-centre
positions $\mathbf{r}\left(t\right)$ at a particular
wave vector, $\mathbf{q}$, taken usually to be the inverse cell size
$q\equiv\left|\mathbf{q}\right|=2\pi/a$. The long-time
decay of $F\left(q,t\right)$ is characterised by the so-called \emph{alpha-relaxation
time} $\tau_{\alpha}$ at which $F\left(q,t\right)$ has decayed by
half. When the system solidifies, i.e. when neighbour exchanges stop, $\mathbf{r}\left(t\right)$ remains constant and hence $\tau_{\alpha}$ diverges,\cite{berthier2011dynamical}
and stays infinite within the solid phase. In Fig.~\ref{fig:phase_diagram}a-c,
we show the phase diagram of $\tau_{\alpha}$ as a function of $p_{0}$
and $f_{\text{a}}$, for several systems with different boundary
conditions. 

In Fig.~\ref{fig:phase_diagram}a we show regions of solid-like and
liquid-like phases in a system with fixed boundaries, at $\Gamma=1$,
and a low noise value of $\tau_{r}=0.01$. We find a boundary of the
solid-like phase that stretches from $p_{0}\approx3.81$ at small
$f_{\text{a}}$ to a maximum activity $f_{\text{a}}$
beyond which the system is fluid at all $p_0$.
This is qualitatively, but not quantitatively consistent with the
results of Bi, \emph{et al.}, who find a transition line at roughly
twice our $f_{\text{a}}$ values. Several factors are likely implicated in this discrepancy. Our systems, at $N=1000$ cells are
more than twice as large as the $N=400$ systems considered by Bi,
\emph{et al.}, and finite system size effects seem to play an important
role, as shown below. We measure $\tau_{\alpha}$ at a value of $1/q$
corresponding to displacements of one cell size. However, even
though displacements are large, we have evidence that this may
not be sufficient to induce T1 transitions and therefore fluidise
the system. Finally, fixed boundaries were used here and the periodic boundaries
of Bi, \emph{et al.} are likely not strictly equivalent.

The influence of the type of boundary conditions is very significant.
In Fig.~\ref{fig:phase_diagram}b, we show the phase diagram for the
same $\Gamma=1$ and $\tau_{r}^{-1}=0.01$ as in Fig.~\ref{fig:phase_diagram}a, except with open boundary
conditions and boundary line tension $\lambda=0$. Separately, for $\Gamma=0.1$,
we have also confirmed that the value of the boundary line tension
does not significantly affect the onset of the solid-like regime (not shown).
We find a significantly
lower maximum $f_{a}$ for the transition, $f_{\text{a}}=0.03$,
a factor of $10$ compared to the fixed case. The effect also persists
at $\tau_{r}^{-1}=0.1$, but is less pronounced (not shown). While we do not have
a full explanation for this result, we do note that fluctuations of
the boundary allow for rearrangements that are otherwise strongly
suppressed by the fixed boundary. For example, the system in Fig.~\ref{fig:gallery}d
shows significant boundary fluctuations. It is liquid-like with $\tau_{\alpha}\approx10$,
whereas the equivalent fixed system has $\tau_{\alpha}\approx100$.
In view of the significant role of the boundary, we expect a strong
system-size dependence.\cite{in-prep} 

At very high $p_{0}$ and low active driving, we observe a systematic
increase of $\tau_{\alpha}$ (especially visible in Fig.~\ref{fig:phase_diagram}c).
This unexpected result is accompanied by structural changes in the
cell patterns that we observe. Fig.~\ref{fig:gallery}f shows a liquid
system at $p_{0}=3.95$, near the relative minimum $\tau_{\alpha}$
for a given $f_{\text{a}}$. The distribution of cell centres
appears random. In contrast, as can be seen in Fig.~\ref{fig:gallery}g, at very high $p_{0}=4.85$, cells arrange
themselves into rosette shapes, where many vertices meet in a point.
Rosettes are a feature of many developmental systems,\cite{fletcher2014vertex}
so it is interesting to see that they do appear naturally in the
AVM context. Cell centres also arrange themselves in equidistant chains,
hinting at a connection to one of the various pattern-formation instabilities studied in nonlinear dynamics.
We note that this regime is numerically delicate, and the addition of
the soft repulsive core between cell centres (see Appendix I) is necessary
to make simulations stable. At present it is not clear if these effects
are artefacts of the AVM, or have real biological significance. 

In parts of the phase diagram, we observe a fingering
instability\cite{poujade2007collective,petitjean2010velocity,lee2011crawling,tarle2015fingering}
where regions a few cells wide migrate outward from the centre, as
shown in Fig.~\ref{fig:gallery}e. When $\tau_{\alpha}$ drops below approximately $10$, we observe
that the fluctuations of the boundary already present
in Fig.~\ref{fig:gallery}d become unconstrained. 
This is a mechanically unstable
regime: Eventually, these cells will detach, a process we are not
yet able to model due to the topological change that it would imply
(see, Fig.~\ref{fig:boundary_topology}). We have observed that fluctuations
need to reach a threshold of approximately $>5\%$ of a length increase
in the boundary to break through to an unconstrained growth, otherwise
the system remains stable, see e.g. Fig.~\ref{fig:gallery}d-g. We then
associate a time scale $\tau_{\text{inst}}$ with reaching this threshold
and use it to measure the degree of instability: a small time scale
denotes a rapid growth rate of fluctuations. 
Fig.~\ref{fig:phase_diagram}c shows $\tau_{\alpha}$, and Fig.~\ref{fig:phase_diagram}d shows $\tau_{\text{inst}}$ for the same open
system with $\Gamma=0.1$, $\tau_{r}^{-1}=0.1$ and boundary line
tension $\lambda=0.1$. We note that the transition line between solid-like
and fluid-like states is low, at $f_{\text{a}}=0.03$. At and below $f_{\text{a}}=0.1$,
the boundary of the system is stable, and above this threshold, the instability becomes
more pronounced with increasing $f_{\text{a}}$ and smaller
$\tau_{\alpha}$. The physical mechanism responsible for the instability
involves a subtle interplay of $f_{\text{a}}$, boundary line
tension (stronger line tension suppresses the instability), the noise
level $\tau_{r}^{-1}$ (lower $\tau_{r}^{-1}$ enhances the instability),
and $p_{0}$. The instability resembles observations of finger formation
in MDCK monolayers.\cite{poujade2007collective,petitjean2010velocity}
Existing models link it to either leader cells,\cite{sepulveda2013collective,tarle2015fingering}
a bending instability,\cite{lee2011crawling} or an active growth
feedback loop,\cite{kopf2013continuum} while here it emerges naturally. A detailed account of this
phenomenon will be published elsewhere.\cite{in-prep}

\subsection{Growth and division}

\label{subsec:growth_and_division}Division and death processes are
important in any living tissue, for example, cell division and ingression processes
play essential roles during development. Therefore, as noted in
Sec. \ref{subsec:grow-div-death}, the AVM is equipped to handle such
processes. It is important to note, however, that the removal of one
cell during apoptosis or ingression and the addition of two new cells
during division in the AVM causes a discrete change in the Voronoi
tesselation which implies a discontinuous change of the local forces
derived from the VM. We have simulated the growth of a small cluster of
cells to assess whether this discrete change in geometry can lead
to any instabilities in the model. These test runs did not reveal
any artefacts due to discontinuities in the force caused by the division
events. 

\begin{figure}[h]
\centering{}\includegraphics[width=0.9\columnwidth]{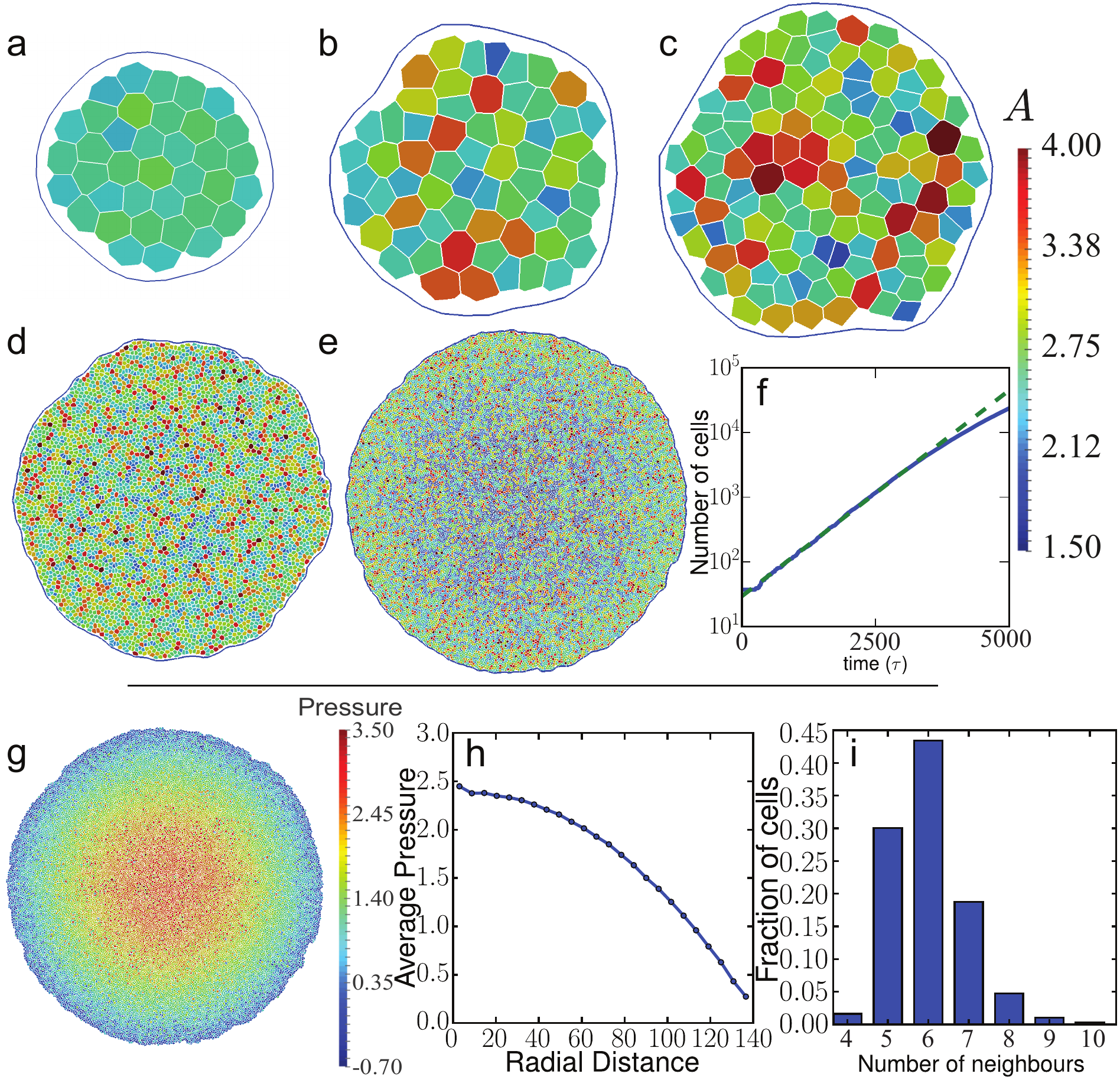}
\caption{Snapshots of a growing epithelial tissue. Frames (a), (b), (c), (d)
and (e) have $37$, $63$, $124$, $4633$ and $23787$ cells and
are at times $50$, $500$, $1000$, $3500$ and $5000$,
respectively. Cells have a chance to divide if their area is greater
than a critical area $A_{c}=2.8a^{2}$ after which the probability
of a cell to divide increases linearly with its area (see Eq.~(\ref{eq:div_prob})).
In this simulation, the shape factor was set to $p_{0}=3.10$. (f)
Log-linear plot of the total number of cells as a function of simulation
time. The growth rate of the patch is initially exponential but starts
to slow at around $3000$ cells. This is due to cells in the centre
of the cluster being prevented from expanding by the surrounding tissue.
(g) Tissue after $5000$ time units ($10^{6}$ time steps) with each cell coloured by pressure.
Pressure has built up in the centre of the tissue while close to the
edge the average pressure is low. (h) Average pressure (averaged over
the polar angle) as a function of the radial distance from the centre
of the tissue. (i) Distribution of the number of neighbours for cells
in the system shown in (g). \label{fig:growth}}
\end{figure}

In order to illustrate the growth process, we choose a shape factor,
$p_{0}=3.10$, corresponding to $\Gamma=1$ and $\Lambda=-5.5$ and
no active driving, i.e., we set $f_{\text{a}}=0$. This puts the system
into the solid-like phase where T1 events are absent. Our simulation
runs for $10^{6}$ time steps at $\delta t = 0.005$, corresponding to $5000$ time units, starting from 37 cells and stopping
at about 24,000 cells. To balance computational efficiency with a
smooth rate of division, cells are checked for division every 25 time
steps. We show snapshots of different stages of the tissue growth
in Fig.~\ref{fig:growth}a-e.

We note that the numerical stability of the simulation that involves
growth is quite sensitive to the values of the parameters used in
the AVM. For example, divisions of highly irregularly shaped cells,
as commonly observed in the high $p_{0}$ regime, can put a significant
strain on the simulation and even cause a crash. Helpfully, some of
these problems can be alleviated with the help of the the soft repulsive
potential defined in Eq.~(\ref{eq:v_soft}) in Appendix I between
cell centres that acts to mediate the impact of cell divisions. Finally,
as a rule, regardless of the exact parameter regime, a smaller time
step is typically required for simulating growing systems. 

In Fig.~\ref{fig:growth}f we show the tissue size as a function of
the simulation time. In this simulation there are no apoptosis or
cell ingression events and, as expected, the tissue size grows exponentially.
However, at long times, the growth slows down and deviates from exponential
growth. This is easy to understand, as the centre of the tissue is
prevented from expanding by the surrounding cells. The effect can
be seen in Fig.~\ref{fig:growth}e, where cells located towards the
centre have, on average, smaller areas and in Fig.~\ref{fig:growth}g,
which shows a clear pressure buildup in the centre. This suggests
that in the later stages, the simulated tissue is not in mechanical equilibrium any more.
The pressure is computed using the
Hardy stress description.\cite{admal2010unified} The details of the
stress calculation in the AVM will be published elsewhere.
We also see clear heterogeneities in the local pressure shown in Fig.~\ref{fig:growth}g.
In Fig.~\ref{fig:growth}h, we show the radial
pressure profile in the tissue at the end of the simulation.  From the
figure it is also evident that the there is a substantial pressure
buildup close to the centre of the tissue as well as that angular
averaging substantially reduces local pressure fluctuation notable
in Fig.~\ref{fig:growth}g. 
The origin of these effects warrants a detailed investigation
and will be addressed in a later publication, 
we note however that stress inhomogeneities are a persistent feature of the epithelial cell monolayers
that have been investigated by traction force microscopy.\cite{trepat2009physical,tambe2011collective}

Finally, in Fig.~\ref{fig:growth}i we
show the distribution of the number of neighbours for this model system.
The observed distribution is in a good agreement with the observations
in actual tissues.\cite{gibson2006emergence,sandersius2011correlating}

\subsection{Modelling mechanically heterogeneous tissues}

\label{subsec:mechanical_heterogeneities}
\begin{figure}[h]
\centering{}\includegraphics[width=0.99\columnwidth]{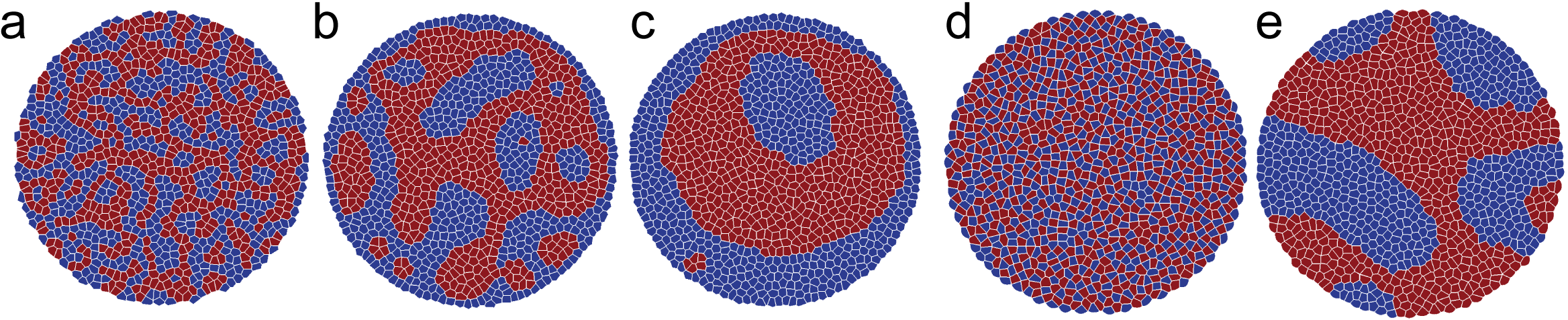} 
\caption{(a-c) Snapshots of a system of two cell types at times $10$, $500$
and $5000$ with $\Lambda_{rb}=-6.4$. (d) A ``checkerboard''
pattern formes immediately (at time $10$) when red-blue cell-cell
contacts are energetically favourable compared with pairs of red-red
and blue-blue contacts, $\Lambda_{rb}=-6.7$. (e) Same initial system
as (a-c) but with red-boundary contacts slightly favoured over blue-boundary
contacts. The system gradually separates into compartments of each
cell type. The uncorrelated random fluctuations are sufficient to
drive neighbour exchanges within the bulk of both the red and blue
cell compartments. Cells on the compartment boundary can sometimes
move parallel to it but meet strong resistance when trying to move across
it. \label{fig:sortims}}
 
\end{figure}

The AVM is equipped to allow for
cell-specific parameters, which enables us to investigate
tissues with locally varying mechanical properties. A commonly studied
example of the effects such heterogeneities is cell sorting. As an example
we show simulations that display sorting of two distinct cell types. We achieve this by setting the junction tension $\Lambda$
for each pair of cell-cell and cell-boundary contacts. All our simulations
consist of $1000$ cells with half chosen randomly to be of the
``red'' type and the others being of the ``blue'' type. In these
simulations, boundaries have been kept fixed. We observe sorting behaviour
akin to that found in other commonly used tissue models.\cite{graner1992simulation,graner1993can2}
Using $r$, $b$ and $M$ to denote red, blue and the boundary, respectively,
we start by fixing $K=1$ and $\Gamma=1$ and set $-6.8=\Lambda_{rr}<\Lambda_{rb}<\Lambda_{bb}=-6.2$,
corresponding to $p_{0}$ in the range $3.58-3.93$. We, however,
note that $p_{0}$ is the quantity defined ``per cell'' and one
should understand it in this context only as a rough estimate whether
a given cell type is in the solid or fluid phase. All cells are subject
to small random fluctuations of their position and polarity vector
$\mathbf{n}_{i}$, which allows for T1 transitions that can bring
initially distant cells into contact. We set the active driving to
zero, i.e., $f_{\text{a}}=0$. We have chosen different values for $\Lambda_{rr}$
and $\Lambda_{bb}$ to reflect the idea that the surfaces of these
cells have different adhesive properties.\cite{foty2005differential}
Note that the $\Lambda$ parameter for a particular contact is proportional
to its energy per unit length. Sorting of cells into groups of the
same type occurs when the energy of two red-blue contacts is greater
than the energy of one red-red contact and one blue-blue contact,
corresponding to $\Lambda_{rb}>\left(\Lambda_{rr}+\Lambda_{bb}\right)/2$,
see Fig.~\ref{fig:sortims}a-c. In this regime, for cells of the same
type it is energetically favourable for the new contact to elongate
while local red-blue contacts are shortened. Conversely, if $\Lambda_{rb}<\left(\Lambda_{rr}+\Lambda_{bb}\right)/2$
then cells maximise their red-blue contacts forming a ``checkerboard''
pattern (Fig.~\ref{fig:sortims}d). The final pattern is not without
defects, the number and location of which depend on the initial conditions.
The tissue boundary consists of contacts between cell centres and
boundary particles so $\Lambda_{rM}$ and $\Lambda_{bM}$ need also
to be specified to reflect the way in which the cell types interact
with the extracellular matrix or surrounding medium. Initially we
set $\Lambda_{rM}=\Lambda_{bM}=-6.2$ and observe that blue cells
cover the boundary enveloping red cells because this facilitates lower
energy red-blue and red-red contacts being formed. If we incentivise
red-boundary contacts by setting $\Lambda_{rM}<\Lambda_{rr}+\Lambda_{bM}-\Lambda_{rb}=-6.6$
we make red-boundary contacts preferable.\cite{graner1993can2} This
case is shown in Fig.~\ref{fig:sortims}e for $\Lambda_{rM}=-6.8$.

\subsection{Effects of cellular alignment}

\label{subsec:cell_cell_align} 
\begin{figure}[h]
\centering{}\includegraphics[width=1\textwidth]{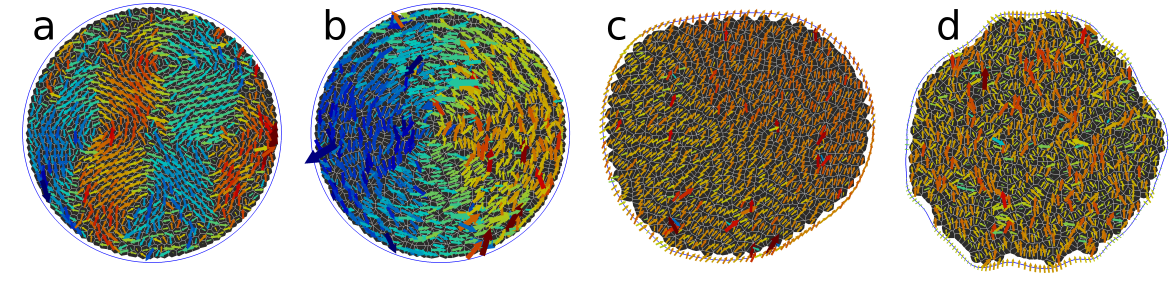}
\caption{Effects of alignment. (a-c) are self-alignment of the cell polarisation
with the cell velocity. (d) is self-alignment of the cell polarisation
with the long axis of the cell. (a) A confined system in the solid-like
region ($f_{\text{a}}=0.03$, $p_{0}=3.385$, $\tau_{r}^{-1}=0.01$),
at alignment strength $J_{v}=1.0$ shows oscillating collective modes.
(b) In the liquid region ($f_{\text{a}}=0.1$, $p_{0}=4.4$,
$\tau_{r}^{-1}=0.1$), the system enters a vortex rotation state instead
($J_{v}=1.0$). (c) An open system at $f_{\text{a}}=0.1$,
$p_{0}=4.4$, $\tau_{r}^{-1}=0.1$, boundary line tension $\lambda=0.1$
and $J_{v}=0.1$ migrates collectively. (d) An open system at $f_{\text{a}}=0.1$,
$p_{0}=3.95$, $\tau_{r}^{-1}=0.1$, boundary line tension $\lambda=0.1$
and alignment with cell shape with $J_{s}=1$ migrates collectively
with complex fluctuations. Arrows are cell velocity vectors $\mathbf{v}_{i}$;
they are coloured by the magnitude of $v_{x}$ for (a) and $v_{y}$
for (b-d). Orange is positive (pointing right/upwards) and blue is
negative (pointing left/downward). \label{fig:alignment} }
\end{figure}

We now briefly turn our attention to the effects of several models
of cell polarity alignment. So far, we have assumed torque $\boldsymbol{\tau}_{i}=0$
in Eq.~(\ref{eq:motion_direction}), i.e., we are in the situation where
the the polarisation vector of each cell is independent of the surrounding
cells and its direction diffuses randomly over time. In biological
systems, it is known that a cell's polarity responds to the surrounding
and many forms of polarity alignment have been proposed. Here,
we highlight two alignment mechanisms that are compatible with the
current understanding of cell mechanics. In Fig.~\ref{fig:alignment}a-c,
we have used the alignment model defined in Eq.~(\ref{eq:velocity_align})
that assumes that the polarity vector $\mathbf{n}_{i}$ of cell $i$
aligns with its velocity $\mathbf{v}_{i}$. The torque term in Eq.
(\ref{eq:motion_direction}) is then given by $\boldsymbol{\tau}_{i}=-J_{v}\mathbf{v}_{i}\times\mathbf{n}_{i}$,
where $J_{v}$ is the alignment strength. This model was first developed
for collectively migrating cells (modelled as particles),\cite{szabo2006phase}
and it exhibits global polar migration, i.e. a state in which all
particles align their velocities and travel as a flock. In dense systems
of active particles confined to a finite region, velocity alignment
has been shown to be intimately linked to collective elastic oscillations.\cite{henkes2011active}
It is remarkable that the main hallmarks of this active matter dynamics are also observed in
the model tissue. In Fig.~\ref{fig:alignment}a, we show velocity
alignment dynamics for a confined system in the solid-like phase;
here the collective oscillations are very apparent. They are
strikingly reminiscent of the collective displacement modes observed
in confined MDCK cell layers.\cite{serra2012mechanical,deforet2014emergence}

In Fig.~\ref{fig:alignment}b, we apply the same dynamics, but now
to a system that is in the liquid-like phase, with fixed boundaries.
Here, the collective migration wins, but the confinement to a disk
with fixed boundaries forces the cells into a vortex-shaped migration
pattern. Finally, in Fig.~\ref{fig:alignment}c, in the absence of
confinement, we recover the collective polar migration of the cell
patch.

In Fig.~\ref{fig:alignment}d, we show the effects of aligning the cell's
polarity to the largest principal axis of the cell shape tensor, defined
in Eq.~(\ref{eq:shape_align}). This type of alignment also leads
to collective motion in an unconfined system, however there are significant
fluctuations as the allowed cell patterns are highly frustrated by
the constraint to remain in a Voronoi tesselation.

These preliminary results serve as a showcase of the non-trivial effects
of cell-cell alignment on the collective behaviour of the entire
tissue. A more detailed account of the effects of different alignment
models will be published elsewhere. 

\subsection{Modelling non-circular tissue shapes}

\begin{figure}
\begin{centering}
\includegraphics[width=0.85\textwidth]{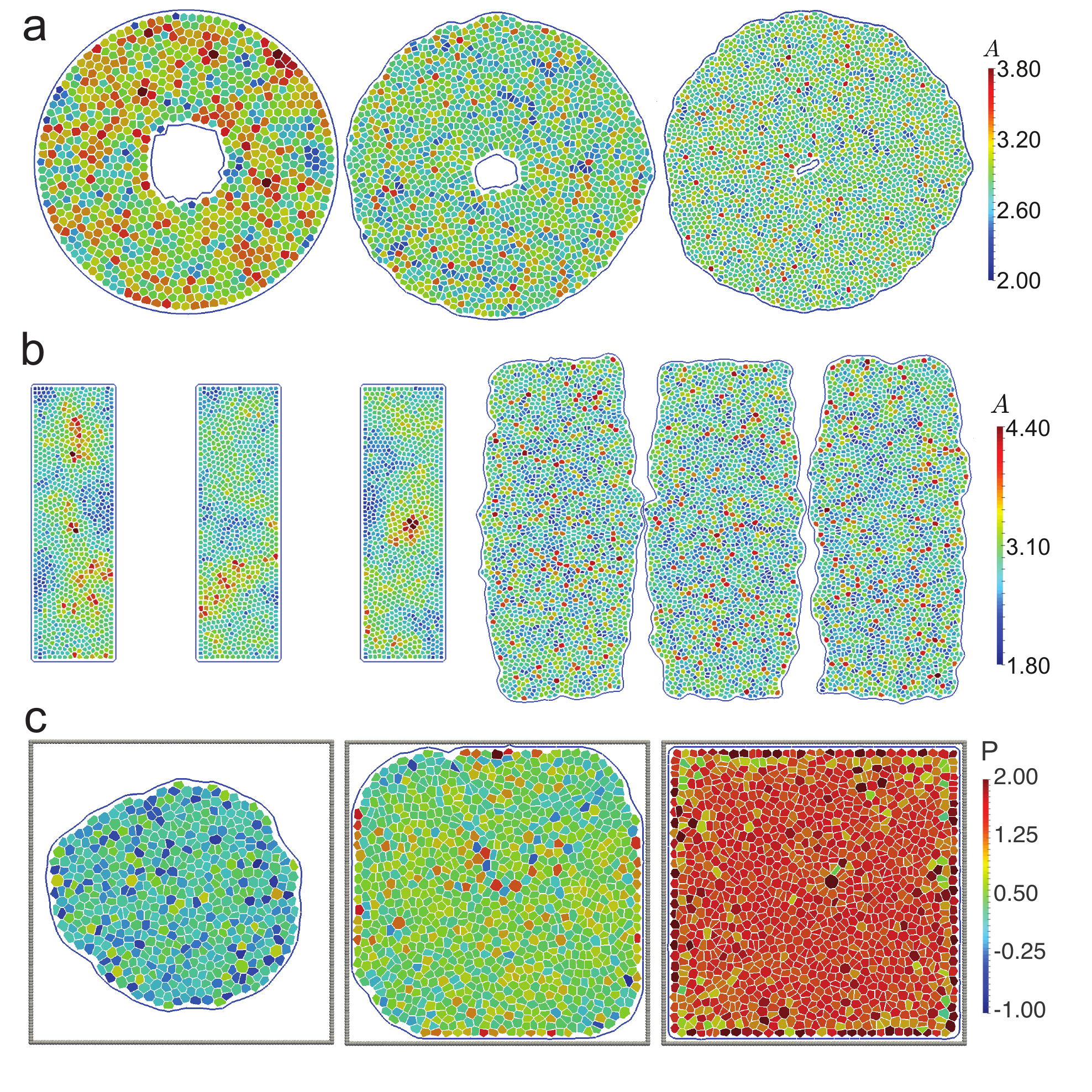}
\par\end{centering}
\caption{(a) Snapshots of the simulation of a cell sheet with an annular geometry
used to illustrate how cells divide and fill the circular void in
the centre. The system configuration was recorded at times $0$,
$1000$ and $1700$. The initial configuration has $p_{0}=3.46$,
i.e., it was in the solid-like phase. While it is not simple to define
$p_{0}$ for a growing system due to a constant change in the cell
target area $A_{0}$, we note that throughout the simulation, cell
shapes remain regular. (b) Illustration of a common experimental
setup where cells are grown in rectangular ``moulds'' for a system
with initial $p_{0}=3.35$. Once the entire region is filled with
cells, the mould is removed and the colony is allowed to freely grow.
Images on the right show two of the strips about to merge. In (c), we model tissue
growth in confinement for a system with initial $p_{0}=3.46$. Grey
beads form the boundary of the confinement region that constrain
the cell growth. Initially, cells do not touch the wall and freely
grow. As the colony reaches the wall, one starts to notice pressure
buildup. Finally, the entire cavity is filled with cells and any subsequent
division leads to increase of the pressure in the system. Snapshots
were recorded at times $1200$, $1480$ and $1600$.
\label{fig:shapes}}
\end{figure}

\label{subsec:shapes}
In the previous discussions, all examples assumed
a patch of cells with the topology of a disk. However, the AVM is not restricted to the
circular geometry and can be applied to systems of arbitrary shapes,
including domains with complex connectivity. Such situations often
arise when modelling experimental systems where cells surround an obstacle,
or in studies of wound healing. In Fig.~\ref{fig:shapes} we present
a gallery of non-circular shapes that can be readily studied using
the AVM as it is implemented in \emph{SAMoS}. The annular geometry shown
in Fig.~\ref{fig:shapes}a would be suited for modelling wound healing
problems as well as situations where cells migrate in order to fill
a void. A common experimental setup where cell colonies are prepared
as rectangular strips\cite{poujade2007collective} is shown in Fig.~\ref{fig:shapes}b, where three
separate patches grow towards each other. Finally, in Fig.~\ref{fig:shapes}c
we show an example of yet another very interesting situation, where
cells are grown in a confined region of space. 

\section{Summary and conclusions}

\label{sec:conclusions}

In this paper we have introduced the Active Vertex Model. It is a hybrid
model that combines ideas from the physics of active matter with the Vertex
Model, a widely used description for modelling confluent epithelial
tissues. Active matter physics is a rapidly growing field of research
in soft condensed matter physics, and it is emerging as a natural
framework for describing many biophysical processes, in particular
those that occur at mesoscales, i.e., at the scales that span multiple
cells to the entire tissue. Our approach is complementary to the recently
introduced Self-Propelled Voronoi model,\cite{bi2016motility} for
it allows modelling of systems with fixed and open, i.e. dynamically
changing boundaries as well as cell-cell alignment, cell growth, division
and death. The AVM has been implemented into the \emph{SAMoS} software
package and is publicly available under a non-restrictive open source
license.

The AVM utilises a mathematical duality between Delaunay and Voronoi
tessellation in order to relate forces on cell centres to the positions
of the vertices of the dual lattice, i.e. meeting points of three
of more cells - a natural description of a confluent epithelial tissue.
This not only allows for a straightforward and efficient implementation
using standard algorithms for particle-based simulations, but provides
a natural framework for modelling topological changes in the tissue,
such as intercalation and ingression events. In other words, in the
AVM T1 transition events arise spontaneously and it is not necessary
to perform any additional steps in order to ensure that cells are
able to exchange neighbours. 

Furthermore, our implementation of the AVM is very efficient, allowing
for simulations of systems containing tens of thousands of cells on
a single CPU core, thus enabling one to probe collective features,
such as global cell flow patterns that span length-scales of several
millimetres. In addition, the AVM is also able to handle multiple
cell types and type specific cell contacts, which allows simulations
of mechanically heterogeneous systems. 

All these features make the AVM a strong candidate model to address
many interesting biological and biophysical problems related to the
mechanical response of epithelial tissues, especially those that occur
at large length and time scales that are typically only accessible
to continuum models. Unlike in the case of continuum models where
relating parameters of the model to the experimental systems is often
difficult and unclear, the AVM retains the cell-level resolution,
making it simpler to connect it to the processes that occur at scales
of single cells. The AVM is, however, not designed to replace continuum
models, but to serve as the important layer that connects the complex
molecular processes that occur at the cell level with the global collective
behaviour observed at the level of the entire tissue. 

With that in mind, there are, of course, still many ways the model
can be improved. For examples, it would be very interesting to augment
the AVM to include the effects of biochemical signalling. This would require
solving a set of differential equations for signals in each time step,
and then supplying those solutions to the mechanics part of the model.
Adding such functionality would substantially increase the computational
cost of simulations, however at the same time it would allow for detailed
studies of the coupling between chemical and mechanical signalling. These
are believed to be essential for developing a full understanding of
the mechanics of epithelial tissues. Given the layout of the AVM and
its implementation, implementing such functionality would be straightforward. 

Furthermore, in the current version of the AVM activity is introduced
in a very rudimentary manner, via assuming that cells self-propel
in the direction of their polarity vector. This is a very strong assumption
that would need a much stronger experimental support than currently
available. It is also possible that a far more sophisticated model
would be required to fully capture cell's motility. However, one also
needs to keep in mind that there is a tradeoff between being as biologically
accurate as possible and retaining a sufficient level of simplicity
to be able to efficiently perform simulations of large systems. With
all this in mind, we argue that the our simulations clearly show that
even this simplistic model of self-propulsion is capable of capturing
many features of real systems, and that it can serve as a good starting
point for building biologically more accurate descriptions. 

Another potentially very interesting feature that is currently not
supported would be splitting and merging of the boundaries, that is,
allowing for topological changes of the entire sheet such as those depicted
in Fig.~\ref{fig:boundary_topology}c. This would us to study detachment of a part of the tissue or opening apertures as well
as the opposite problem of closing holes and gaps. The latter is of
great importance for studying problems related to wound healing. Unfortunately,
setting up a set of general rules on how to automatically split a
boundary line or merge two boundaries into a single one is not a simple
task from a point of view of computational geometry. The problem is
further complicated if those rules are also to be made biologically
plausible, which is essential for the model to be relevant to actual
experiments.

We conclude by noting that extending the model to a curved surface
or making it fully three-dimensional would be more involved. Being
able to study curved epithelial sheets, however, would be of great
interest to systems where curvature clearly cannot be ignored, e.g.,
as in the case in modelling intestinal crypts.\cite{clevers2013intestinal}
While there is nothing in the description of the AVM that is unique
to the planar geometry, there are several technical challenges associated
with directly porting it onto a curved surface. Most notably, building
a Delaunay triangulation on an arbitrary curved surface is not a simple
task. In addition, quantities such as bending rigidity that are naturally
defined on triangles would have to be properly mapped onto contacts
lines between neighbouring cells. This is not straightforward
to do. Developing a fully three-dimensional version of the AVM would
be an even a greater challenge since the duality between Delaunay and Voronoi
descriptions central to this model has no analogue in three dimensions. 

We hope that the AVM will provide a useful and complementary
tool for probing the aspects of the epithelial tissue mechanics that
are not available to other methods, as well as serve as an independent
validation for the results obtained by other methods.

\section{Acknowledgements }

We would like to acknowledge many valuable discussions with Geoff
Barton, Dapeng (Max) Bi, Luke Coburn, Amit Das, Tamal Das, Alexander
Fletcher, M Lisa Manning, M Cristina Marchetti, Kirsten Martens, Inke
N\"athke, and K R Prathyusha. SH acknowledges hospitality of the Kavli
Institute of Theoretical Physics during the workshop ``From Genes
to Growth and Form''. RS and CJW acknowledge support by the UK BBRSC
(grant BB/N009789/1) and SH acknowledges support by the UK BBSRC (grant
BB/N009150/1).

\section*{Appendix I: Force on the cell centre}

In this appendix we derive the expression for forces on the cell centre,
Eq.~(\ref{eq:force_expression}), starting from the expression for
the VM energy given in Eq.~(\ref{eq:vertex_model}). The force on
cell $i$ is computed as the negative derivative with respect to $\mathbf{r}_{i}$
of the energy functional $E_{VM}$ in Eq.~(\ref{eq:vertex_model}).
Conceptually, this is a straightforward calculation with the only
real complication being that $E_{VM}$ is most naturally written in
terms of the positions of the Voronoi vertices, $\mathbf{r}_{\mu}$, while
in the AVM we track positions of cell centres. In general, 
\begin{equation}
\mathbf{F}_{i}=-\nabla_{\mathbf{r}_{i}}E_{VM}.\label{eq:general_force_expression}
\end{equation}
When computing the gradient in the last expression, we need to keep
in mind that moving particle $i$ changes the shapes of all of its neighbouring
cells. Therefore, moving cell $i$ exerts a force on a number of cell
centres in its surrounding. All those contributions have to be taken
into account when computing the force $\mathbf{F}_{i}$. A direct consequence
of this coupling between the cell and all of its neighbouring cells
is that the force $\mathbf{F}_{i}$ \emph{cannot} be written as a
simple sum of pairwise interactions between cell $i$ and each of
its neighbours. We'll get back to this point below.

We start by computing the direct contributions resulting from moving particle
$i$ itself. The area term will produce a force 
\begin{eqnarray}
\mathbf{F}_{i}^{area} & = & -\nabla_{\mathbf{r}_{i}}\frac{K_{i}}{2}\left(A_{i}-A_{i}^{0}\right)^{2}=-K_{i}\left(A_{i}-A_{i}^{0}\right)\nabla_{\mathbf{r}_{i}}A_{i}.\label{eq:force_area_step_1}
\end{eqnarray}
We need to compute $\nabla_{\mathbf{r}_{i}}A_{i}\equiv\frac{\partial A_{i}}{\partial\mathbf{r}_{i}}$,
which we write by using the vector form of the chain rule for the
calculating derivatives, 
\begin{equation}
\left[\nabla_{\mathbf{r}_{i}}A_{i}\right]_{k}=\sum_{\nu\in\Omega_{i}}\left[\nabla_{\mathbf{r}_{\nu}}A_{i}\right]_{m}\left[\frac{\partial\mathbf{r}_{\nu}}{\partial\mathbf{r}_{i}}\right]_{mk},\label{eq:area_gradient}
\end{equation}
where the sum is over all vertices surrounding particle $i$, referred
to as the \emph{loop} $\Omega_{i}$ of particle $i$. $\left[\frac{\partial\mathbf{r}_{\nu}}{\partial\mathbf{r}_{i}}\right]$
is the $3\times3$ Jacobian matrix of the transformation between coordinates
of cell centres and positions of the vertices of the dual Voronoi
tessellation. $\left[\dots\right]_{k}$ represents the $k^{\text{th}}-$component of
the gradient vector, with $k\in\left\{ x,y,z\right\} $ and we assume
summation over the repeated index $m$. The Jacobian can be computed
using the barycentric coordinates that connect centres and vertices introduced in Eq.~(\ref{eq:r_mu_baricentric}) and Fig.~\ref{fig:model}. We have 
\begin{equation}
\frac{d\mathbf{r}_{\mu}}{d\mathbf{r}_{p}}=\mathbf{r}_{i}\otimes\frac{d}{d\mathbf{r}_{p}}\left(\frac{\lambda_{1}}{\Lambda}\right)+\frac{\lambda_{1}}{\Lambda}\delta_{ip}\hat{I}+\mathbf{r}_{j}\otimes\frac{d}{d\mathbf{r}_{p}}\left(\frac{\lambda_{2}}{\Lambda}\right)+\frac{\lambda_{2}}{\Lambda}\delta_{jp}\hat{I}+\mathbf{r}_{k}\otimes\frac{d}{d\mathbf{r}_{p}}\left(\frac{\lambda_{3}}{\Lambda}\right)+\frac{\lambda_{3}}{\Lambda}\delta_{kp}\hat{I}.\label{eq:rmu_rp}
\end{equation}
where
\begin{align}
\lambda_{1} & =l_{i}^{2}\left(l_{j}^{2}+l_{k}^{2}-l_{i}^{2}\right)\nonumber \\
\lambda_{2} & =l_{j}^{2}\left(l_{k}^{2}+l_{i}^{2}-l_{j}^{2}\right)\nonumber \\
\lambda_{3} & =l_{k}^{2}\left(l_{i}^{2}+l_{j}^{2}-l_{k}^{2}\right),\label{eq:circum_lambda}
\end{align}
are the barycentric coordinates and $\Lambda=\lambda_{1}+\lambda_{2}+\lambda_{3}$
with 
\begin{align}
l_{i} & =\left|\mathbf{r}_{j}-\mathbf{r}_{k}\right|=\left|\mathbf{r}_{jk}\right|\nonumber \\
l_{j} & =\left|\mathbf{r}_{k}-\mathbf{r}_{i}\right|=\left|\mathbf{r}_{ki}\right|\nonumber \\
l_{k} & =\left|\mathbf{r}_{i}-\mathbf{r}_{j}\right|=\left|\mathbf{r}_{ij}\right|.\label{eq:circum_l}
\end{align}
$\hat{I}$ is a $3\times3$ identity matrix, and $\otimes$ stands
for the outer product between two vectors. We proceed by calculating
\begin{equation}
\frac{d}{d\mathbf{r}_{p}}\left(\frac{\lambda_{q}}{\Lambda}\right)=\frac{1}{\Lambda^{2}}\left(\Lambda\frac{d\lambda_{q}}{d\mathbf{r}_{p}}-\lambda_{q}\frac{d\Lambda}{d\mathbf{r}_{p}}\right),\label{eq:lambda_q_Lambda}
\end{equation}
for $q=1,2,3$. It is straightforward to show that 
\begin{align}
\frac{d\lambda_{q}}{d\mathbf{r}_{p}} & =\left(L^{2}-4l_{q}^{2}\right)\frac{d\left(l_{q}^{2}\right)}{d\mathbf{r}_{p}}+l_{q}^{2}\frac{d\left(L^{2}\right)}{d\mathbf{r}_{p}},\label{eq:lambda_q_rp}
\end{align}
where $l_{q}$ is defined in Eq.~(\ref{eq:circum_l}) and $L^{2}=l_{1}^{2}+l_{2}^{2}+l_{3}^{2}$.
We readily calculate, 
\begin{align}
\frac{d\left(l_{i}^{2}\right)}{d\mathbf{r}_{p}} & =\begin{cases}
0 & \,\,\mathrm{for}\,\,p=i\\
2\mathbf{r}_{jk} & \,\,\mathrm{for}\,\,p=j\\
-2\mathbf{r}_{jk} & \,\,\mathrm{for}\,\,p=k
\end{cases},\label{eq:li2_rp}
\end{align}
\begin{equation}
\frac{d\left(l_{j}^{2}\right)}{d\mathbf{r}_{p}}=\begin{cases}
-2\mathbf{r}_{ki} & \,\,\mathrm{for}\,\,p=i\\
0 & \,\,\mathrm{for}\,\,p=j\\
2\mathbf{r}_{ki} & \,\,\mathrm{for}\,\,p=k
\end{cases},\label{eq:lj2_rp}
\end{equation}
\begin{equation}
\frac{d\left(l_{k}^{2}\right)}{d\mathbf{r}_{p}}=\begin{cases}
2\mathbf{r}_{ij} & \,\,\mathrm{for}\,\,p=i\\
-2\mathbf{r}_{ij} & \,\,\mathrm{for}\,\,p=j\\
0 & \,\,\mathrm{for}\,\,p=k
\end{cases},\label{eq:lk2_rp}
\end{equation}
where $\mathbf{r}_{ij}=\mathbf{r}_{i}-\mathbf{r}_{j}$, etc. Combining
the last three expressions gives 
\begin{equation}
\frac{d\left(L^{2}\right)}{d\mathbf{r}_{p}}=\begin{cases}
2\left(-\mathbf{r}_{ki}+\mathbf{r}_{ij}\right) & \,\,\mathrm{for}\,\,p=i\\
2\left(\mathbf{r}_{jk}-\mathbf{r}_{ij}\right) & \,\,\mathrm{for}\,\,p=j\\
2\left(-\mathbf{r}_{jk}+\mathbf{r}_{ki}\right) & \,\,\mathrm{for}\,\,p=k
\end{cases}.\label{eq:L2_rp}
\end{equation}
In order to reduce the computational effort, it is convenient to precompute
and store derivatives for all three values of $p=i,j,k$. We start
with the case $p=i$ for which we have, 
\begin{equation}
\frac{d\lambda_{1}}{d\mathbf{r}_{i}}=2\left|\mathbf{r}_{jk}\right|^{2}\left(-\mathbf{r}_{ki}+\mathbf{r}_{ij}\right).\label{eq:lambda_1_i}
\end{equation}
\begin{align}
\frac{d\lambda_{2}}{d\mathbf{r}_{i}} & =-2\left(\left|\mathbf{r}_{jk}\right|^{2}+\left|\mathbf{r}_{ij}\right|^{2}-2\left|\mathbf{r}_{ki}\right|^{2}\right)\mathbf{r}_{ki}+2\left|\mathbf{r}_{ki}\right|^{2}\mathbf{r}_{ij}.\label{eq:lambda_2_i}
\end{align}
\begin{equation}
\frac{d\lambda_{3}}{d\mathbf{r}_{i}}=2\left(\left|\mathbf{r}_{jk}\right|^{2}+\left|\mathbf{r}_{ki}\right|^{2}-2\left|\mathbf{r}_{ij}\right|^{2}\right)\mathbf{r}_{ij}-2\left|\mathbf{r}_{ij}\right|^{2}\mathbf{r}_{ki}.\label{eq:lambda_3_i}
\end{equation}
For $p=j$ we have, 
\begin{align}
\frac{d\lambda_{1}}{d\mathbf{r}_{j}} & =2\left(\left|\mathbf{r}_{ki}\right|^{2}+\left|\mathbf{r}_{ij}\right|^{2}-2\left|\mathbf{r}_{jk}\right|^{2}\right)\mathbf{r}_{jk}-2\left|\mathbf{r}_{jk}\right|^{2}\mathbf{r}_{ij}.\label{eq:lambda_1_j}
\end{align}
\begin{align}
\frac{d\lambda_{2}}{d\mathbf{r}_{j}} & =2\left|\mathbf{r}_{ki}\right|^{2}\left(\mathbf{r}_{jk}-\mathbf{r}_{ij}\right).\label{eq:lambda_2_j}
\end{align}
\begin{align}
\frac{d\lambda_{3}}{d\mathbf{r}_{j}} & =-2\left(\left|\mathbf{r}_{jk}\right|^{2}+\left|\mathbf{r}_{ki}\right|^{2}-2\left|\mathbf{r}_{ij}\right|^{2}\right)\mathbf{r}_{ij}+2\left|\mathbf{r}_{ij}\right|^{2}\mathbf{r}_{jk}.\label{eq:lambda_3_j}
\end{align}
For $p=k$ we have, 
\begin{align}
\frac{d\lambda_{1}}{d\mathbf{r}_{k}} & =-2\left(\left|\mathbf{r}_{ki}\right|^{2}+\left|\mathbf{r}_{ij}\right|^{2}-2\left|\mathbf{r}_{jk}\right|^{2}\right)\mathbf{r}_{jk}+2\left|\mathbf{r}_{jk}\right|^{2}\mathbf{r}_{ki}.\label{eq:lambda_1_k}
\end{align}
\begin{align}
\frac{d\lambda_{2}}{d\mathbf{r}_{k}} & =2\left(\left|\mathbf{r}_{jk}\right|^{2}+\left|\mathbf{r}_{ij}\right|^{2}-2\left|\mathbf{r}_{ki}\right|^{2}\right)\mathbf{r}_{ki}-2\left|\mathbf{r}_{ki}\right|^{2}\mathbf{r}_{jk}.\label{eq:lambda_2_k}
\end{align}
\begin{equation}
\frac{d\lambda_{3}}{d\mathbf{r}_{k}}=2\left|\mathbf{r}_{ij}\right|^{2}\left(-\mathbf{r}_{jk}+\mathbf{r}_{ki}\right).\label{eq:lambda_3_k}
\end{equation}
Finally, 
\begin{align}
\frac{d\Lambda}{d\mathbf{r}_{i}} & =-4\left(\left|\mathbf{r}_{jk}\right|^{2}+\left|\mathbf{r}_{ij}\right|^{2}-\left|\mathbf{r}_{ki}\right|^{2}\right)\mathbf{r}_{ki}+4\left(\left|\mathbf{r}_{jk}\right|^{2}+\left|\mathbf{r}_{ki}\right|^{2}-\left|\mathbf{r}_{ij}\right|^{2}\right)\mathbf{r}_{ij}.\label{eq:LAMBDA_i}
\end{align}
\begin{align}
\frac{d\Lambda}{d\mathbf{r}_{j}} & =4\left(\left|\mathbf{r}_{ki}\right|^{2}+\left|\mathbf{r}_{ij}\right|^{2}-\left|\mathbf{r}_{jk}\right|^{2}\right)\mathbf{r}_{jk}-4\left(\left|\mathbf{r}_{jk}\right|^{2}+\left|\mathbf{r}_{ki}\right|^{2}-\left|\mathbf{r}_{ij}\right|^{2}\right)\mathbf{r}_{ij}.\label{eq:LAMBDA_j}
\end{align}
\begin{align}
\frac{d\Lambda}{d\mathbf{r}_{k}} & =-4\left(\left|\mathbf{r}_{ki}\right|^{2}+\left|\mathbf{r}_{ij}\right|^{2}-\left|\mathbf{r}_{jk}\right|^{2}\right)\mathbf{r}_{jk}+4\left(\left|\mathbf{r}_{jk}\right|^{2}+\left|\mathbf{r}_{ij}\right|^{2}-\left|\mathbf{r}_{ki}\right|^{2}\right)\mathbf{r}_{ki}.\label{eq:LAMBDA_k}
\end{align}
These expressions allow us to compute all derivatives in Eq.~(\ref{eq:lambda_q_Lambda})
and, in turn, the Jacobian in Eq.~(\ref{eq:rmu_rp}).

We still need to compute the derivative of the cell's area with
respect to the positions of the vertices of the Voronoi cell. A straightforward
calculation starting from Eq.~(\ref{eq:area_particle}) gives

\begin{eqnarray}
\left[\frac{\partial A_{i}}{\partial\mathbf{r}_{\nu}}\right]_{k} & = & \left[\frac{\partial}{\partial\mathbf{r}_{\nu}}\left\{ \frac{1}{2}\sum_{\mu\in\Omega_{i}}\left(\mathbf{r}_{\mu}\times\mathbf{r}_{\mu+1}\right)\cdot\mathbf{N}_{i}\right\} \right]_{k}\nonumber \\
 & = & \left[\frac{1}{2}\frac{\partial}{\partial\mathbf{r}_{\nu}}\sum_{\mu\in\Omega_{i}}\varepsilon_{\alpha\beta\gamma}x_{\beta}^{\left(\mu\right)}x_{\gamma}^{\left(\mu+1\right)}N_{\alpha}^{\left(i\right)}\right]_{k}\nonumber \\
 & = & \frac{1}{2}\sum_{\mu\in\Omega_{i}}\varepsilon_{\alpha\beta\gamma}N_{\alpha}^{\left(i\right)}\frac{\partial}{\partial x_{k}^{\left(\nu\right)}}\left(x_{\beta}^{\left(\mu\right)}x_{\gamma}^{\left(\mu+1\right)}\right)\nonumber \\
 & = & \frac{1}{2}\varepsilon_{\alpha k\gamma}N_{\alpha}^{\left(i\right)}x_{\gamma}^{\left(\nu+1\right)}+\frac{1}{2}\varepsilon_{\alpha\beta k}N_{\alpha}^{\left(i\right)}x_{\beta}^{\left(\nu-1\right)}\nonumber \\
 & = & \frac{1}{2}\varepsilon_{k\gamma\alpha}N_{\alpha}^{\left(i\right)}x_{\gamma}^{\left(\nu+1\right)}-\frac{1}{2}\varepsilon_{k\beta\alpha}N_{\alpha}^{\left(i\right)}x_{\beta}^{\left(\nu-1\right)}\nonumber \\
 & = & \frac{1}{2}\left[\mathbf{r}_{\nu+1}\times\mathbf{N}_{i}-\mathbf{r}_{\nu-1}\times\mathbf{N}_{i}\right]_{k},\label{eq:area_rnu}
\end{eqnarray}
where $\varepsilon_{\alpha\beta\gamma}$ is the Levi-Chivita symbol.
Therefore, the expression for the area change due to displacing vertex $i$
is 
\begin{equation}
\left[\nabla_{\mathbf{r}_{i}}A_{i}\right]_{k}=\frac{1}{2}\sum_{\nu\in\Omega_{i}}\left[\mathbf{r}_{\nu+1}\times\mathbf{N}_{i}-\mathbf{r}_{\nu-1}\times\mathbf{N}_{i}\right]_{m}\left[\frac{\partial\mathbf{r}_{\nu}}{\partial\mathbf{r}_{i}}\right]_{mk},\label{eq:area_ri}
\end{equation}
where, as above, we have assumed summation over the repeated index $m$.
We finally derive the force on vertex $i$ due to the area contractions,
\begin{equation}
\mathbf{F}_{i}^{area}=-\frac{1}{2}K_{i}\left(A_{i}-A_{i}^{0}\right)\sum_{\nu\in\Omega_{i}}\left[\mathbf{r}_{\nu+1}\times\mathbf{N}_{i}-\mathbf{r}_{\nu-1}\times\mathbf{N}_{i}\right]^{T}\left[\frac{\partial\mathbf{r}_{\nu}}{\partial\mathbf{r}_{i}}\right],\label{eq:force_area}
\end{equation}
where $T$ in superscript stands for transpose, i.e., $\left[\dots\right]^{T}\left[\dots\right]$
stands for a matrix product between a $1\times3$ matrix (i.e., a
vector) and $3\times3$ Jacobian matrix.

We can now proceed to calculate derivatives of the second term in
Eq.~(\ref{eq:vertex_model}). The perimeter of cell $i$ is defined as
\begin{equation}
P_{i}=\sum_{\mu\in\Omega_{i}}\left|\mathbf{r}_{\mu+1}-\mathbf{r}_{\mu}\right|.\label{eq:perim}
\end{equation}
As above, we calculate 
\begin{eqnarray}
\left[\frac{\partial P_{i}}{\partial\mathbf{r}_{\nu}}\right]_{k} & = & \left[\frac{\partial}{\partial\mathbf{r}_{\nu}}\sum_{\mu\in\Omega_{i}}\left|\mathbf{r}_{\mu+1}-\mathbf{r}_{\mu}\right|\right]_{k}\nonumber \\
 & = & \sum_{\mu\in\Omega_{i}}\frac{\partial}{\partial x_{k}^{\left(\nu\right)}}\left[\sum_{\alpha}\left(x_{\alpha}^{\left(\mu+1\right)}-x_{\alpha}^{\left(\mu\right)}\right)^{2}\right]^{1/2}\nonumber \\
 & = & \sum_{\mu\in\Omega_{i}}\frac{1}{\left|\mathbf{r}_{\mu+1}-\mathbf{r}_{\mu}\right|}\sum_{\alpha}\left(x_{\alpha}^{\left(\mu+1\right)}-x_{\alpha}^{\left(\mu\right)}\right)\left(\delta_{\alpha k}\delta_{\nu,\mu+1}-\delta_{\alpha k}\delta_{\nu,\mu}\right)\nonumber \\
 & = & \frac{x_{k}^{\left(\nu\right)}-x_{k}^{\left(\nu-1\right)}}{\left|\mathbf{r}_{\nu}-\mathbf{r}_{\nu-1}\right|}-\frac{x_{k}^{\left(\nu+1\right)}-x_{k}^{\left(\nu\right)}}{\left|\mathbf{r}_{\nu+1}-\mathbf{r}_{\nu}\right|}.\label{eq:perim_rnu}
\end{eqnarray}
We therefore have 
\begin{eqnarray}
\left[\nabla_{\mathbf{r}_{i}}P_{i}\right]_{k} & = & \sum_{\nu\in\Omega_{i}}\left[\frac{x_{m}^{\left(\nu\right)}-x_{m}^{\left(\nu-1\right)}}{\left|\mathbf{r}_{\nu}-\mathbf{r}_{\nu-1}\right|}-\frac{x_{m}^{\left(\nu+1\right)}-x_{m}^{\left(\nu\right)}}{\left|\mathbf{r}_{\nu+1}-\mathbf{r}_{\nu}\right|}\right]_{m}\left[\frac{\partial\mathbf{r}_{\nu}}{\partial\mathbf{r}_{i}}\right]_{mk},\label{eq:perim_ri}
\end{eqnarray}
where we also assume summation over the repeated index $m$. The force
term resulting from perimeter contractions is then given as 
\begin{equation}
\mathbf{F}_{i}^{perim}=-\Gamma_{i}P_{i}\sum_{\nu\in l_{i}}\left[\hat{\mathbf{r}}_{\nu,\nu-1}-\hat{\mathbf{r}}_{\nu+1,\nu}\right]^{T}\left[\frac{\partial\mathbf{r}_{\nu}}{\partial\mathbf{r}_{i}}\right],\label{eq:force_perim}
\end{equation}
where we have defined $\hat{\mathbf{r}}_{\nu,\nu-1}=\frac{\mathbf{r}_{\nu}-\mathbf{r}_{\nu-1}}{\left|\mathbf{r}_{\nu}-\mathbf{r}_{\nu-1}\right|}$,
etc.

A similar calculation for the the last term in Eq.~(\ref{eq:vertex_model})
leads to 
\[
\nabla_{\mathbf{r}_{i}}l_{\nu,\nu+1}=\sum_{\nu\in l_{i}}\left[\frac{\mathbf{r}_{\nu}-\mathbf{r}_{\nu-1}}{\left|\mathbf{r}_{\nu}-\mathbf{r}_{\nu-1}\right|}-\frac{\mathbf{r}_{\nu+1}-\mathbf{r}_{\nu}}{\left|\mathbf{r}_{\nu+1}-\mathbf{r}_{\nu}\right|}\right]^{T}\left[\frac{\partial\mathbf{r}_{\nu}}{\partial\mathbf{r}_{i}}\right],
\]
where we have explicitly labelled the two nearest neighbours (in the counterclockwise
direction as) as $\nu$ and $\nu+1$. The force due to cell junction
contractions is then computed as

\begin{equation}
\mathbf{F}_{i}^{junct}=-\sum_{\nu\in l_{i}}\left[\Lambda_{\nu-1,\nu}\frac{\mathbf{r}_{\nu}-\mathbf{r}_{\nu-1}}{\left|\mathbf{r}_{\nu}-\mathbf{r}_{\nu-1}\right|}-\Lambda_{\nu,\nu+1}\frac{\mathbf{r}_{\nu+1}-\mathbf{r}_{\nu}}{\left|\mathbf{r}_{\nu+1}-\mathbf{r}_{\nu}\right|}\right]^{T}\left[\frac{\partial\mathbf{r}_{\nu}}{\partial\mathbf{r}_{i}}\right].\label{eq:force_junction}
\end{equation}

We now move to the second part, which is to determine the force on
particle $i$ as a result of displacing one of its surrounding particles.
In order to do this, we go back to the original expression for the
energy in the VM, Eq.~(\ref{eq:vertex_model}). The force on vertex $i$
is then 
\begin{eqnarray*}
\mathbf{F}_{i} & = & -\nabla_{\mathbf{r}_{i}}E_{VM}\\
 & = & -\nabla_{\mathbf{r}_{i}}\left\{ \sum_{k=1}^{N}\left[\frac{K_{k}}{2}\left(A_{k}-A_{k}^{0}\right)^{2}+\frac{\Gamma_{k}}{2}P_{k}^{2}\right]+\sum_{\left\langle \mu,\nu\right\rangle }\Lambda_{\mu\nu}l_{\mu\nu}\right\} ,
\end{eqnarray*}
where $N$ is the total number of cells. We first focus on the area
and perimeter terms. We have 
\begin{eqnarray*}
\mathbf{F}_{i}^{a+p} & = & -\nabla_{\mathbf{r}_{i}}\sum_{k=1}^{N}\left[\frac{K_{k}}{2}\left(A_{k}-A_{k}^{0}\right)^{2}+\frac{\Gamma_{k}}{2}P_{k}^{2}\right]\\
 & = & -\sum_{k=1}^{N_{part}}\sum_{\nu\in\Omega_{k}}\left(\nabla_{\mathbf{r}_{\nu}}\left[\frac{K_{k}}{2}\left(A_{k}-A_{k}^{0}\right)^{2}+\frac{\Gamma_{k}}{2}P_{k}^{2}\right]\right)^{T}\left[\frac{\partial\mathbf{r}_{\nu}}{\partial\mathbf{r}_{i}}\right].
\end{eqnarray*}
Further, we have 
\begin{eqnarray}
\mathbf{F}_{i}^{a+p} & = & -\sum_{k=1}^{N}\sum_{\nu\in\Omega_{k}}\left[K_{k}\left(A_{k}-A_{k}^{0}\right)\left(\nabla_{\mathbf{r}_{\nu}}A_{k}\right)+\Gamma_{k}P_{k}\left(\nabla_{\mathbf{r}_{\nu}}P_{k}\right)\right]^{T}\left[\frac{\partial\mathbf{r}_{\nu}}{\partial\mathbf{r}_{i}}\right]\nonumber \\
 & = & -\sum_{k=1}^{N}\frac{K_{k}}{2}\left(A_{k}-A_{k}^{0}\right)\sum_{\nu\in\Omega_{k}}\left[\left(\mathbf{r}_{\nu+1}-\mathbf{r}_{\nu-1}\right)\times\mathbf{N}_{k}\right]^{T}\left[\frac{\partial\mathbf{r}_{\nu}}{\partial\mathbf{r}_{i}}\right]\nonumber \\
 &  & -\sum_{k=1}^{N}\Gamma_{k}P_{k}\sum_{\nu\in\Omega_{k}}\left(\frac{\mathbf{r}_{\nu}-\mathbf{r}_{\nu-1}}{\left|\mathbf{r}_{\nu}-\mathbf{r}_{\nu-1}\right|}-\frac{\mathbf{r}_{\nu+1}-\mathbf{r}_{\nu}}{\left|\mathbf{r}_{\nu+1}-\mathbf{r}_{\nu}\right|}\right)^{T}\left[\frac{\partial\mathbf{r}_{\nu}}{\partial\mathbf{r}_{i}}\right].\label{eq:area_perim_force_total}
\end{eqnarray}
From the last expression it is clear that only vertices that are displaced
by moving cell $i$ are going to contribute to the force. These vertices
are all ``corners'' of cell $i$ and a subset of ``corners'' of
its immediate neighbours affected by $i$. This gives us the algorithm
for computing the total force on particle $i$ coming from the area: 
\begin{enumerate}
\item Loop over particle $i$ and all its neighbours.
\begin{enumerate}
\item For particle $i$ compute $\frac{K_{i}}{2}\left(A_{i}-A_{i}^{0}\right)$
and multiply it by the sum $\sum_{\nu\in\Omega_{k}}\left[\left(\mathbf{r}_{\nu+1}-\mathbf{r}_{\nu-1}\right)\times\mathbf{N}_{i}\right]^{T}\left[\frac{\partial\mathbf{r}_{\nu}}{\partial\mathbf{r}_{i}}\right]$.
Note that this sum is over all vertices (corners) $\nu$ of the cell
$i$. 
\item For all immediate neighbours $j$ of cell $i$ compute $\frac{K_{j}}{2}\left(A_{j}-A_{j}^{0}\right)$
and multiply it with the sum $\sum_{\nu\in l_{i}\cap l_{j}}\left[\left(\mathbf{r}_{\nu+1}-\mathbf{r}_{\nu-1}\right)\times\mathbf{N}_{j}\right]^{T}\left[\frac{\partial\mathbf{r}_{\nu}}{\partial\mathbf{r}_{i}}\right]$.
Note that $\nu\in l_{i}\cap l_{j}$ ensures that vertices $\nu$ surrounding
$j$ are taken into account only if they are affected by (and also
belong to) cell $i$.  
\end{enumerate}
\end{enumerate}
A similar algorithm can be used to compute force contribution of the
perimeter term.

We now focus on the last term, which is the force along the cell junctions.
If we note that we can write the cell junction term as 
\[
E^{j}=\sum_{\mu}\Lambda_{\mu,\mu+1}l_{\mu,\mu+1},
\]
we have 
\begin{eqnarray}
\mathbf{F}_{i}^{j} & = & -\sum_{k=1}^{N}\sum_{\nu\in\Omega_{k}}\left[\Lambda_{\nu-1,\nu}\frac{\mathbf{r}_{\nu}-\mathbf{r}_{\nu-1}}{\left|\mathbf{r}_{\nu}-\mathbf{r}_{\nu-1}\right|}-\Lambda_{\nu,\nu+1}\frac{\mathbf{r}_{\nu+1}-\mathbf{r}_{\nu}}{\left|\mathbf{r}_{\nu+1}-\mathbf{r}_{\nu}\right|}\right]^{T}\left[\frac{\partial\mathbf{r}_{\nu}}{\partial\mathbf{r}_{i}}\right].\label{eq:force_junction_total}
\end{eqnarray}
As before, we loop over all vertices that are affected by changing
the position of cell $i$, which are all corners of the cell $i$ and
a subset of corners of its immediate neighbours whose positions are
determined by $\mathbf{r}_{i}$.

Combining Eqs. (\ref{eq:area_perim_force_total}) and (\ref{eq:force_junction_total})
then ultimately leads to Eq.~(\ref{eq:force_expression}).  We note that in Eq.~(\ref{eq:force_junction_total}) we have 
absorbed factor of $1/2$ arising from double counting edges into the definition of $\Lambda_{\nu,\nu+1}$.

Finally, we briefly address another contribution to the force which
needs to be added to the model allow for simulations deep in the liquid-like
phase. As discussed in Sec. \ref{subsec:phase_diagram}, similar to
the SPV model, the AVM shows a transition between solid-like and fluid-like
phases. In the solid-like phase, intercalation events are inhibited
and without cell division and death, cells do not exchange their neighbours.
In the fluid-like phase, on the other hand, cells are much more mobile
and one observes a large number of T1 transitions and intercalations.
In this regime, the Delaunay triangulation is very irregular with
many obtuse triangles. If a triangle becomes very obtuse, its circumcenter
is far outside the triangle and even small changes in the position
of one of its vertices leads to large movements of the circumcenter.
For any simulation time step that is not extremely small, this can
lead to unphysical self intersections of the triangulation and cause
the simulation to be unreliable, or, in the most extreme cases, causes
it to crash. In practice, in order to prevent this from happening
while still being able to use a reasonably large time step, we endow
each cell centre with a soft repulsive core of radius $a$. The repulsive
potential between neighbouring cell centres is then given as 
\begin{equation}
V_{soft}\left(r_{ij}\right)=\begin{cases}
\frac{1}{2}k\left(r_{ij}-2a\right)^{2} & \,\,\mathrm{for}\,\,r_{ij}<2a\\
0 & \mathrm{otherwise}
\end{cases},\label{eq:v_soft}
\end{equation}
where $r_{ij}=\left|\mathbf{r}_{i}-\mathbf{r}_{j}\right|$. This core
prevents two cell centres from getting too close to each other. The corresponding
force on cell $i$ is 
\begin{equation}
\mathbf{F}_{i}^{soft}=-\nabla_{\mathbf{r}_{i}}V_{soft}\left(r_{ij}\right)=\begin{cases}
-k\left(r_{ij}-2a\right)\hat{\mathbf{r}}_{ij} & \,\,\mathrm{for}\,\,r_{ij}<2a\\
0 & \mathrm{otherwise}
\end{cases}\label{eq:force_soft}
\end{equation}
with $\hat{\mathbf{r}}_{ij}\equiv\left|\mathbf{r}_{i}-\mathbf{r}_{j}\right|/r_{ij}$
and according to the Newton's third law, $\mathbf{F}_{j}^{soft}=-\mathbf{F}_{i}^{soft}$.
The soft repulsion does not interfere with the AVM dynamics except for its regularising effect. In Fig.~\ref{fig:gallery}e-g, the pale spheres drawn at the cell centres have radius $a$, and they remain far removed from the cell boundaries.

We note that this repulsive force is similar in spirit to the limits
that have to be imposed on edge lengths in order to prevent unphysical
self-intersections in the triangulated models for lipid membranes,
that have been extensively studied in the 1990s.\cite{gommper2004incollection} It
is interesting to note that in the case of triangulated models for
lipid membranes one also needs to introduce a maximum allowed edge
length in order to prevent unphysical configurations. In the case
of AVM, this is not necessary.

\section*{Appendix II: Algorithm for handling boundaries}

Due to the dynamic nature of the model, even without cell division
and death, it is not possible to retain a constant number of boundary
particles. Instead, the boundary line has to be able to contract or
extend in order to accommodate changes inside the tissue. This is
achieved by dynamically adding and removing boundary particles.

\begin{figure*}
\begin{centering}
\includegraphics[width=0.9\textwidth]{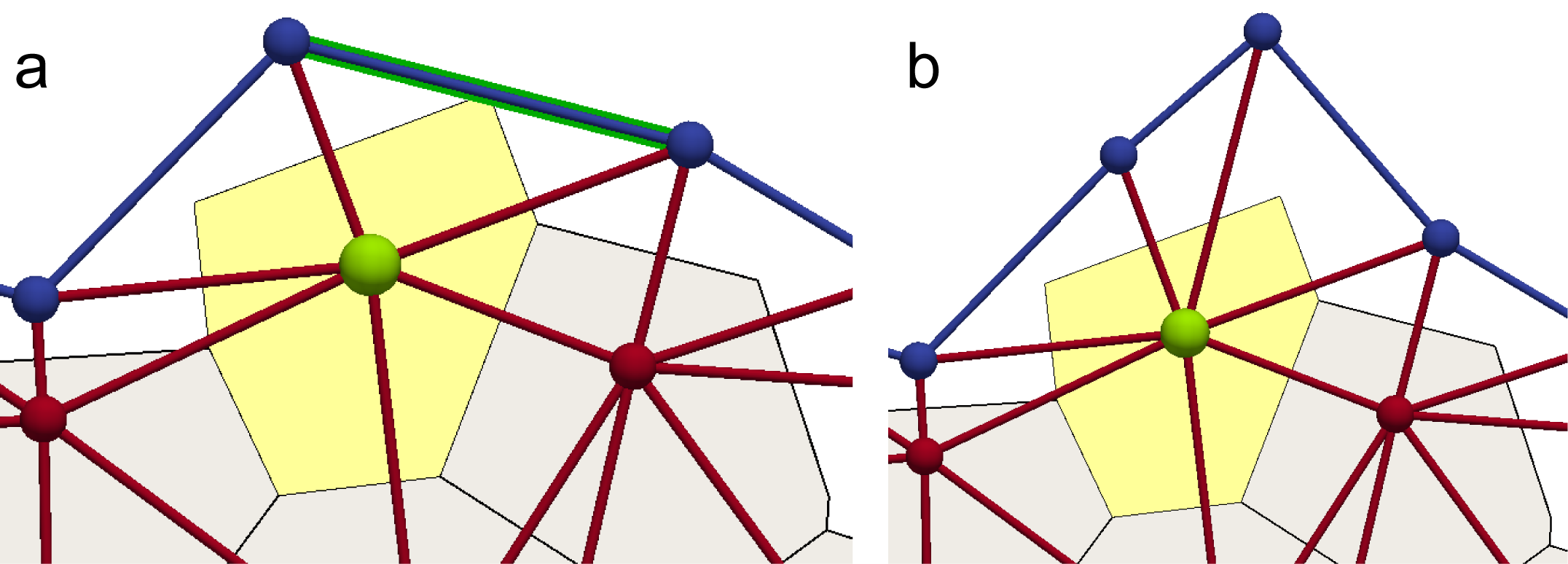} 
\par\end{centering}
\caption{Expansion of the boundary by adding a new boundary particle. (a) If
the angle at the internal particle shaded in green opposite to the
highlighted edge reaches $90^{\circ}$, one of the corners of the
cell (yellow polygon) touches the highlighted edge. This triggers
a ``flip'' mechanism. (b) The internal particle shaded in green
is mirrored along the shaded edge in (a) and a new boundary particle
(top blue) is introduced. The shaded edge is flipped such that the
new particle is connected to the ``green'' one.\label{fig:boundary_vertex_add}}
\end{figure*}

We first focus on the boundary expansion. We require that all cells
are contained within the boundary, that is, no dual vertices belonging
to a cell are allowed to ``spill'' over the boundary line. This
condition is violated if the angle opposite to a boundary edge is
greater than $90^{\circ}$. In this case, the centre of the circumscribed
circle falls outside the triangle and, therefore, outside the boundary.
In order to prevent this from happening we perform the following check
(see also Fig.~\ref{fig:boundary_vertex_add}): 
\begin{enumerate}
\item For each boundary edge $e$ compute angle $\alpha_{e}$ at the particle
$p_{e}$ opposite to it. 
\item If $\alpha>90^{\circ}$ 
\begin{enumerate}
\item Compute the position, $\mathbf{r}_{p_{n}}$, of the new particle $p_{n}$
by mirroring the coordinates of $p_{e}$, $\mathbf{r}_{p_{e}}$, with
respect to edge $e$. If $\text{\ensuremath{\mathbf{\hat{r}}}}_{e}$
is the unit-length vector along edge $e$ then 
\begin{equation}
\mathbf{r}_{p_{n}}=2\left(\mathbf{r}_{p_{e}}\cdot\hat{\mathbf{r}}_{e}\right)\hat{\mathbf{r}}_{e}-\mathbf{r}_{p_{e}}.\label{eq:rpn}
\end{equation}
\item Add a new boundary particle $p_{n}$ at position $\mathbf{r}_{p_{n}}$
and mark it as boundary. 
\item Remove boundary edge $e$, i.e., the two boundary particles at its
end are no longer neighbours. 
\item Connect $p_{n}$ to the two boundary particles disconnected in (c). 
\item Connect $p_{n}$ to $p_{e}$. 
\end{enumerate}
\item If at least one new boundary particle was added in 2., rebuild the
triangulation. 
\end{enumerate}
Note that the procedure outlined above always converges in a single
step.

\begin{figure*}
\begin{centering}
\includegraphics[width=0.9\textwidth]{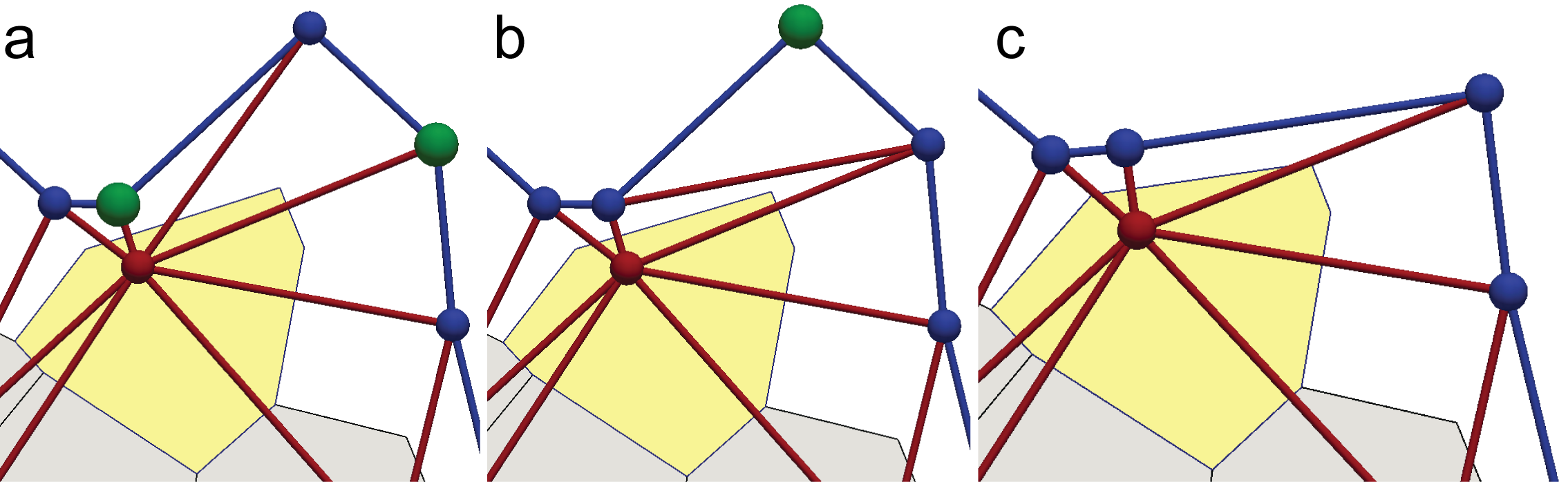} 
\par\end{centering}
\caption{Shrinking of the boundary is achieved by removing boundary vertices
that have only two bonds. (a) If the sum of angles at two particles
shaded in green opposite the edge connecting the internal and boundary
particles is greater than $180^{\circ}$ the edge is flipped (this
is the standard equiangulation move that occurs for all internal edges).
(b) After the flip, the boundary particles shaded in green has only
two bonds, both to its boundary neighbours. (c) The ``green'' particle
in (b) is then removed. \label{fig:boudnary_removal}}
\end{figure*}

Shrinking of the boundary is achieved by removing particles that have
no connections to the internal particles. In this situation, no part
of a cell can be inside a triangle that has two of its sides being
boundary edges and it can be safely removed. The algorithm schematically
outlined in Fig.~\ref{fig:boudnary_removal} is as follows: 
\begin{enumerate}
\item For each boundary particle $p$ compute number of edges $n_{e}\left(p\right)$
that the particle belongs to. 
\item If $n_{e}\left(p\right)\le2$ remove $p$. 
\end{enumerate}
Note that the position of the ``dangling'' particles does not directly
affect the shape of the cell.

An important point to make here is that the algorithms used to add
and remove boundary particles \emph{do not} lead to sudden discontinuous
changes of the shape of the cell or the force acting on its centre.
However, addition and removal of boundary particles inevitably leads
to discontinuous changes in the forces acting on boundary particles.
A potential way to avoid such discontinuous behaviour would be to,
e.g., smoothly ``turn on'' the interactions with newly added particles
or by slowly ``fade out'' interaction with particles that are to
be removed. In practice, however, the discontinuous causes by simply
adding or removing boundary particles lead to changes in the force
that are small and do not appreciably affect the simulation. 

\section*{Appendix III: Implementation}

The AVM is implemented into the \emph{SAMoS} code developed by this
team.\cite{samos2016} In this appendix we first provide a general
overview of the organisation of the \emph{SAMoS} code and then discuss
how the AVM in implemented in it.

\subsection*{SAMoS overview}

\emph{SAMoS} is a software package developed for simulating agent-based
active matter systems confined to move on curved or flat surfaces.\cite{sknepnek2015active}
The code is written in C++, using the C++98 standard with extensive
use of the Standard Template Library and boost libraries.\cite{boost}
It utilises a modular, object oriented design making it very flexible
and simple to extend. It adheres to modern software design principles
and uses a cross-platform build system (\emph{cmake}) as well as automatic
documentation generation with \emph{Doxygen}.\cite{vanHeesch2016doxygen}

\emph{SAMoS} consists of several components each implemented as class
hierarchies. 
\begin{itemize}
\item \textbf{System} - central component that handles the system configuration; 
\item \textbf{Parser} - a recursive descent parser for parsing files that
control execution of the simulation (configuration files). It is implemented
using boost's Phoenix library;\cite{boost2016phoenix} 
\item \textbf{Messenger} - logs system messages (warnings, errors, etc.)
as well as meta-data (parameter trees in JSON or XML format) for data
curation; 
\item \textbf{Neighbour list} - handles build and update of the Verlet neighbour
list;\cite{allen1989computer} 
\item \textbf{Constraint} - handles projection on various flat or curved
surfaces. It is possible to have multiple constraints acting on different
groups of particles; 
\item \textbf{Interactions} - handles all interactions on and between particles.
Interaction parameters can be type-specific, i.e., it is possible
to simulate multicomponent systems. It is also possible to have multiple
interaction types simultaneously present in the system; 
\begin{itemize}
\item \textbf{Pair/multi-body} interactions - handles all interactions that
involve pairs or multiplets of particles (Vertex, Lennard-Jones, soft
repulsion, Morse, etc.); 
\item \textbf{External} interactions - handles all forces that act on a
single particle, such as external fields and/or activity; 
\item \textbf{Bond/angle} interactions - handles interaction between connected
beads for simulating filaments; 
\end{itemize}
\item \textbf{Alignment} - handles alignment of the particle orientation 
\begin{itemize}
\item \textbf{Pair} alignment - handles various models for alignment to
the direction of neighbouring particles (discussed in Sec. \ref{subsec:alignment}); 
\item \textbf{External} alignment - handles alignment to internal or external
cues, such as velocity, cell shape or external fields (discussed in
Sec. \ref{subsec:alignment}); 
\end{itemize}
\item \textbf{Dump} - handles output of the system snapshot in various formats
(VTP, raw text data file, input configuration for restarts, etc.).
It is possible to have multiple dump types present in the same simulation; 
\item \textbf{Log} - handles output of the current system state, such as
total potential energy, mean velocity, temperature, etc.; 
\item \textbf{Population} - handles addition and removal of particles, such
as during cell division and death. It is possible to have multiple
population controls acting at the same time or on different groups
of particles; 
\item \textbf{Integrator} - handles various numerical integrators (Langevin,
Brownian, NVE, etc.) for solving the equations of motion. Different
integrators can acts on different groups of particles (e.g. no motion,
for keeping a subset of particles fixed). 
\end{itemize}
The components are designed to be as loosely coupled as possible,
in order to ensure flexibility, ease of testing and debugging; and
also to make extensions of the code, such as adding a new interaction
force or population control mechanism, as simple as implementing a
new subclass with a minimal need to modify the core of the code.

In order to perform a simulation with \emph{SAMoS}, the user has to
supply two or three text files: 1) a file, referred to as the \emph{data}
file, containing the initial configuration of the system (i.e the
initial positions and velocities of the particles, particle types,
polarity, etc.), 2) for AVM simulations only, a file, referred to
as the \emph{boundary} file, containing the initial labels and connectivity
of the boundary particles and 3) the parameter file, referred to as
the \emph{configuration} file, which sets the simulation protocol
(interaction types and parameters, constraints, type and frequency
of dumps, simulation time step, etc.). Commands in the configuration
file are parsed and executed in the order they appear. Examples of
data and configuration files can be found in the \emph{configurations}
directory in the \emph{SAMoS} installation.

In the current implementation, \emph{SAMoS} runs on a single CPU core.

\subsection*{AVM implementation}

\begin{figure*}
\begin{centering}
\includegraphics[width=0.95\textwidth]{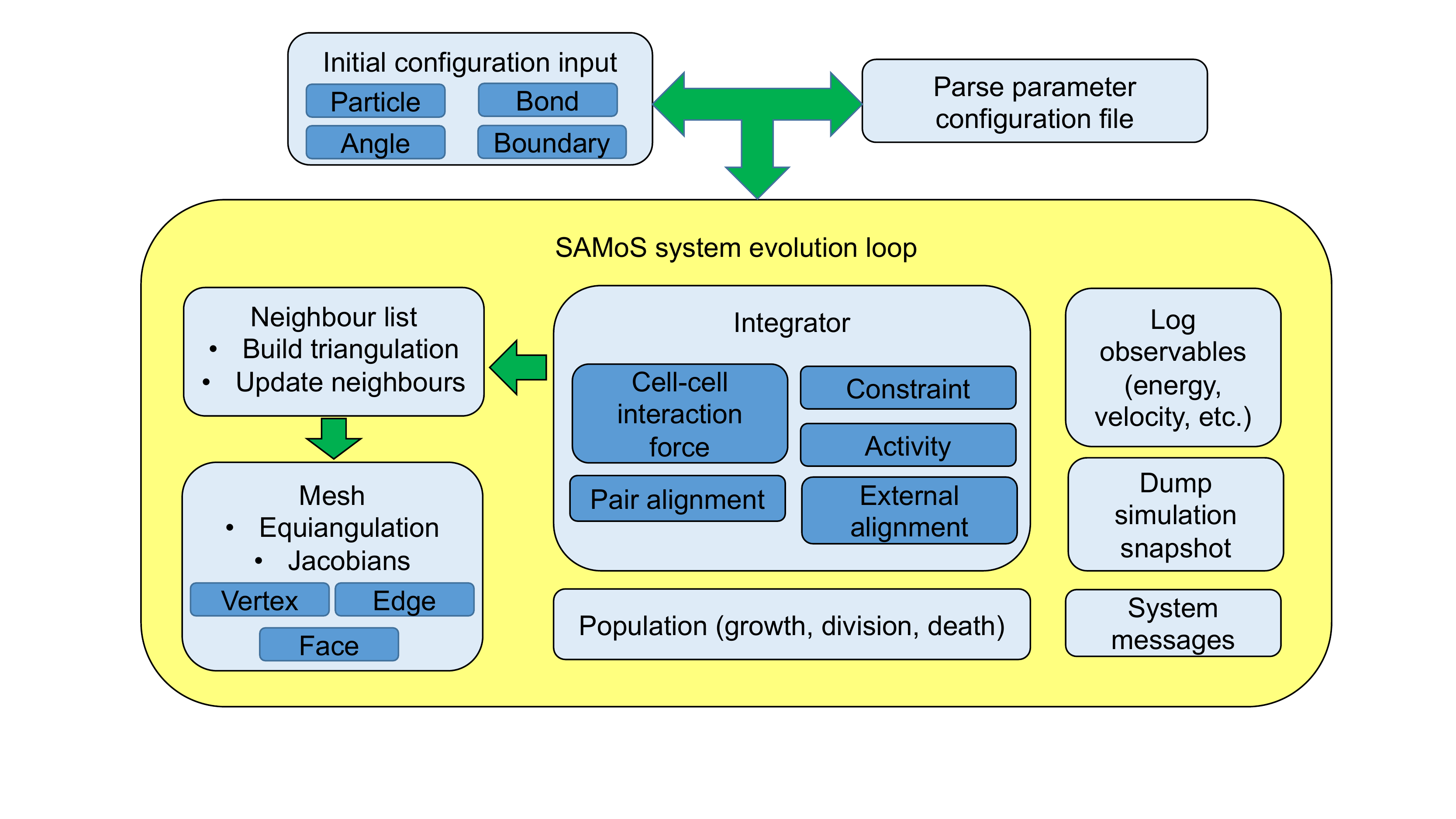} 
\par\end{centering}
\caption{Overview of the general organisation of the AVM implementation into
\emph{SAMoS}.\label{fig:Organisation-of-SAMoS}}
\end{figure*}

The AVM is implemented as an extension of the \emph{SAMoS} code. The
main addition to the code involves a light-weight implementation of
the half-edge data structure\cite{campagna1998directed} as a separate,
\emph{Mesh }class. This class holds the information about the Delaunay
triangulation and computes its dual Voronoi diagram. The \emph{Mesh}
class also ensures that the Delaunay character of the triangulation
is preserved between rebuilds using the equiangulation procedure discussed
in Sec. \ref{subsec:equiangulation}. Finally, \emph{Mesh} computes
the Jacobian matrix, $\left[\frac{\partial\mathbf{r}_{\nu}}{\partial\mathbf{r}_{i}}\right]$
and supplies its elements to the part of the code that computes the
force on the cells. A feedback loop from the integrator ensures that
the \emph{Mesh} class always has the correct position of the cell
centres.

The Delaunay triangulation of the initial positions of the cell centres
is performed using CGAL's Delaunay library.\cite{cgal:eb-16b} In
order to properly include the boundary of the cell sheet, we compute
a constrained Delaunay triangulation,\cite{chew1989constrained} where
the boundary line, supplied by the user as an input file, acts as
the constraint placed on the triangulation. It is important to note
that all Delaunay triangulation libraries always produce a convex
hull of the region that needs to be triangulated. The existence of
the boundary line and the use of the constrained triangulation method
allows us to simply and clearly distinguish between the triangles
that are ``inside'' the tissue and should be kept and the ``outside''
ones that are an artefact of the triangulation procedure and should
be discarded. This weeding out of outside triangles imposes some performance
burden on the computation and we do not perform it at every time step,
but maintain it to be Delaunay by applying equiangulation moves
that are fast to compute. The triangulation is completely rebuilt
only after the steps that involve cell division/death events or boundary
extension/contraction. In order not to disturb the connectivity of
the boundary, only the internal edges can be flipped. In practice,
depending on the parameter values, the triangulation is rebuilt once
every 10-50 time steps. This substantially improves the performance
of the code.

The force on the cell centre, Eq.~(\ref{eq:force_expression}), is
computed by a subclass of the pair/multi-body submodule in the Interactions
module discussed in the previous section. This class invokes the \emph{Mesh}
class in order to obtain the information about positions of dual vertices
and components of the Jacobian matrix.

An overview of the general organisation of the AVM implementation
is shown in Fig.~\ref{fig:Organisation-of-SAMoS}.

 \bibliographystyle{apsrev4-1}
%

\end{document}